# Evidence for a Monolayer Excitonic Insulator

Yanyu Jia[1,#], Pengjie Wang[1,#], Cheng-Li Chiu[1,#], Zhida Song[1], Guo Yu[1,2], Berthold Jäck[1], Shiming Lei[3], Sebastian Klemenz[3], F. Alexandre Cevallos[3], Michael Onyszczak[1], Nadezhda Fishchenko[1], Xiaomeng Liu[1], Gelareh Farahi[1], Fang Xie[1], Yuanfeng Xu[4], Kenji Watanabe[5], Takashi Taniguchi[6], B. Andrei Bernevig[1], Robert J. Cava[3], Leslie M. Schoop[3], Ali Yazdani[1,*], Sanfeng Wu[1,*]

[1] Department of Physics, Princeton University, Princeton, New Jersey 08544, USA
[2] Department of Electrical Engineering, Princeton University, Princeton, New Jersey 08544, USA
[3] Department of Chemistry, Princeton University, Princeton, New Jersey 08544, USA
[4] Max Planck Institute of Microstructure Physics, 06120 Halle, Germany
[5] Research Center for Functional Materials,
National Institute for Materials Science, 1-1 Namiki, Tsukuba 305-0044, Japan
[6] International Center for Materials Nanoarchitectonics,
National Institute for Materials Science, 1-1 Namiki, Tsukuba 305-0044, Japan
[#]These authors contributed equally to this work.
[*] Email: sanfengw@princeton.edu; yazdani@princeton.edu

**Abstract**

The interplay between topology and correlations can generate a variety of quantum phases, many of which remain to be explored. Recent advances have identified monolayer $WTe_2$ as a promising material for doing so in a highly tunable fashion. The ground state of this two-dimensional (2D) crystal can be electrostatically tuned from a quantum spin Hall insulator (QSHI) to a superconductor. However, much remains unknown about the gap-opening mechanism of the insulating state. Here we report evidence that the QSHI is also an excitonic insulator (EI), arising from the spontaneous formation of electron-hole bound states – excitons. We reveal the presence of an intrinsic insulating state at the charge neutrality point (CNP) in clean samples and confirm the correlated nature of this charge-neutral insulator by tunneling spectroscopy. We provide evidence against alternative scenarios of a band insulator or a localized insulator and support the existence of an EI phase in the clean limit. These observations lay the foundation for understanding a new class of correlated insulators with nontrivial topology and identify monolayer $WTe_2$ as a promising candidate for exploring quantum phases of ground-state excitons.

**Main Text**

Different classes of novel quantum phases, including Chern insulators and QSHI, can be generated by various mass acquisition mechanisms of 2D gapless Dirac fermions[1]. A novel mass generation channel in a 2D Dirac semimetal, such as graphene, is the spontaneous exciton formation at charge neutrality[2–5], which in principle can drive a semimetal-insulator phase transition at low temperatures. While the observation of this excitonic gap in graphene has been long-sought after[5], the real Coulomb interaction is so weak that it can only perturbatively reshape the Dirac cones rather than produce an insulating gap[6]. Despite continued theoretical interests[7–11] and despite having potential in engineering exotic states harnessing topology and fractionalization[9,12,13], the experimental observation of a correlated excitonic insulator (EI)[14–21] has not moved beyond engineering quantum Hall bilayers[22–24].

Monolayer WTe$_2$ is a highly unusual 2D crystal, where topology, correlations and spin-orbit coupling (SOC) are simultaneously important[25–30]. Without SOC and interactions, a pair of tilted gapless Dirac points develop at the two sides of the $\Gamma$ point along the $\Gamma$-$X$ direction in the monolayer's Brillouin zone (supplementary section 1), accompanied by a band inversion[25,31]. At low temperatures, the monolayer undergoes a semimetal-insulator transition, entering the QSHI state[25–28]. Fundamental aspects remain unknown, including the gap-opening mechanism. Various possibilities have been suggested. Existing band structure calculations in the literature produce inconsistent results, i.e., either a band insulator phase[27,32,33] (Fig. 1a) or a semimetal phase[25] (Fig. 1b). In the band insulator scenario, a single particle gap is opened via SOC[25–27,32]. A recent scanning tunneling microscope (STM) experiment contradicted this picture and proposed that the insulating state may arise due to a disorder-induced Coulomb gap[34] that diminishes the density of state at Fermi surface. This is an extrinsic mechanism due to electron localization (Fig. 1c). Another possibility is the spontaneous formation of mobile excitons, resulting in an EI phase (Fig. 1d, see supplementary section 1 for our theoretical modeling of this phase and also a recent proposal for relevant compounds[7]). Experimental evidence to uncover the true nature of the charge-neutral ground state in the clean limit of the pristine monolayer is so far lacking. In this work, by combining transport and tunneling measurements, we uncover novel properties of the WTe$_2$ monolayer that support the presence of an EI phase in clean samples.

To reveal transport properties of the bulk of monolayer WTe$_2$, we carefully design our devices so as to avoid transport along the conducting edge channels[26,28] (Fig. 2a and b). A hexagonal boron nitride (hBN) layer with patterned holes is used to cover the metal electrodes except their very ends, which are exposed to WTe$_2$. The electrical contact to the monolayer bulk is then achieved without touching the edges. Top and bottom gates with hBN dielectric are used to tune both the carrier densities and the electric displacement field. Device details and the fabrication procedures are described in the supplementary materials (supplementary section 2). Fig. 2c plots the measured four-probe resistance $R_{xx}$ as a function of the top ($V_{tg}$) and bottom ($V_{bg}$) gate voltages, taken from device D1 at 70 K. The resistance exhibits a sharp peak at the charge neutrality line in the map, where the densities of electrons and holes are equal (see supplementary section 3 and 4 for gate-dependent densities). Below 70 K, the peak resistance in the device is too large to be reliably measured. In Fig. 2d, $R_{xx}$ is plotted as a function of gate-induced carrier density, $n_g \equiv \varepsilon_r\varepsilon_0(V_{tg}/d_{tg}+V_{bg}/d_{bg})/e$, where $e$ is the elementary charge, $\varepsilon_0$ is the vacuum permittivity, $\varepsilon_r$ is the relative dielectric constant of hBN and $d_{tg}$ ($d_{bg}$) is the thickness of hBN layer associated with the top (bottom) gate. When the monolayer is doped with either electrons or holes, $R_{xx}$ drops quickly from its maximum value (> 1 M$\Omega$ at 70 K) at the charge neutrality point (CNP), before a metal-insulator transition occurs (Fig. 2d inset). These observations are consistent with existing reports[26,28–30] on the monolayer, except that the edge state contributions are eliminated and the device quality here is significantly improved, as evidenced by the sharp CNP peak.

The exceptionally high quality of device D1 is achieved, because we used a double graphite gate geometry[35], a flux-grown WTe$_2$ bulk crystal with a large residual-resistance ratio ($RRR$ ~

2500, see supplementary section 5), and a fabrication procedure that minimizes disorder. To investigate the effect of disorder on the insulating state, we fabricated additional devices with controlled quality by altering these conditions, e.g., using a vapor-grown WTe$_2$ bulk[36] with a typical $RRR$ < 400 (D2) or an air-sensitive metal ZrTe$_2$ as the top gate (D3). The size and contact geometries of the three devices are similar. Gate-tuned resistance maps of D2 and D3 are plotted in supplementary section 6. A direct comparison of the three devices is shown in Fig. 2d, where $R_{xx}(n_g)$ curves at 70 K are plotted together. Compared to D1, the peak of D2 or D3 is much broader and displays a significant offset from zero $n_g$, signifying the effect of disorder and inhomogeneity. One may treat the narrowness of the peak as an indicator of the sample cleanness and this quantity is improved by an order of magnitude in D1. In supplementary section 7, we plot the same data in terms of conductivity, which reveals that higher mobility is achieved for both electrons and holes in D1. All these observations confirm that D1 is much cleaner than D2 and D3; yet the resistivity of D1 at the CNP is much higher (Fig. 2d). The contrast experiments hence imply that the insulating state at CNP is an intrinsic property of the monolayer in the clean limit. The disorder-induced Coulomb gap, as discussed previously[34], is thus unsatisfactory to explain our observations.

This intrinsic charge neutral insulator exhibits a strong dependence on the electric displacement field, defined as $D = (V_{bg}/d_{bg} - V_{tg}/d_{tg})\varepsilon_r/2$. The temperature ($T$) dependent $R_{xx}$ typically yields two regimes, separated by a temperature of nearly 100 K, as seen in the Arrhenius plot (Fig. 2e). Figure 2f plots the $T$ dependent $R_{xx}$ of D1 at selected points along the charge neutrality line in Fig. 2c, where $D$ is varied. While the effect of $D$ on the high-$T$ range is not significant, the curve at low-$T$ is clearly flattened by the application of a small $D$. Note that we present the data down to the lowest temperature below which the four-probe measurements become unreliable due to the huge resistance. Due to the lack of a well-established model for EI, in supplementary section 8, we analyze the data based on both the standard activation formula expected for a conventional band insulator and the Efros–Shklovskii (ES) variable-range hopping formula expected for a Coulomb gap[37]. Neither formula describes the observed resistance within the decade of the temperature variation, as already reflected by the presence of the two regimes. In the scenario of a SOC gap, the observation of suppressed $R_{xx}$ at low $T$ would imply closing of the single-particle gap by $D$, which may be reasonable as $D$ breaks inversion symmetry and introduces spin splitting at the band edges[25,33]. However, we will experimentally rule out the band insulator scenario below. Our first-principal band structure calculation shows a semi-metallic band structure prior to the exciton formation. It also shows that $D$ not only introduces spin splitting but also enlarges the Fermi pockets of both electrons and holes simultaneously at charge neutrality (supplementary section 1). To produce a fully gapped EI phase in WTe$_2$ monolayer, one must gap out three Fermi pockets (two electron- and one hole- pockets), which is different from the EI formation in a simple two-band system with one-electron and one-hole pockets. Our Hartree-Fock calculation shows that a full EI gap can indeed develop in the pristine monolayer WTe$_2$ (supplementary section 1). Under the application of $D$, the enlarged Fermi surfaces are expected to reduce the excitonic gap at specific spots in the Brillouin zone. For large $D$, small residual Fermi pockets may be created, coexisting with excitonic order. At low $T$, the residual pockets, if exists, are expected to be localized.

In principle, an EI may be distinguished from other insulators, such as band insulators or Mott insulators, by its Hall response in magnetic fields. In conventional insulators, the vanishing conductivity tensor when lowering $T$ corresponds to a diverging behavior in both $R_{xx}$ and Hall resistances ($R_{xy}$). However, while the diverging $R_{xx}$ is generic for an insulator, $R_{xy}$ can deviate from such an expectation. An example is the so-called "Hall insulator" observed at the vicinity of the quantum Hall liquid state, where $R_{xx}$ diverges yet $R_{xy}$ remains finite[38,39]. An EI is another exception, which can be understood intuitively by considering an idealized case assuming that electrons and holes are symmetric. The formation of excitons will lead to a diverging $R_{xx}$. Yet $R_{xy}$ in this idealized insulator is nevertheless strictly zero at all temperatures, because the contributions from electrons and holes cancel precisely. In real materials, electrons and holes may be asymmetric, e.g., they have different masses or mobilities, then a non-zero $R_{xy}$ can develop. However, the opposite contributions from electrons and holes in the correlated system will still produce anomalous behaviors[40], such as the insensitivity to the insulating resistivity over a wide range of doping and $T$. Despite recent progress[14–20] in identifying an EI using spectroscopic approaches, as far as we know its intrinsic transport properties, including the Hall response, remain elusive. For instance, the prime EI candidate $TiSe_2$ is typically a semimetal even at low $T$ [41,42], preventing transport access to the insulating gap. Our monolayer $WTe_2$ devices are well suited for pursuing such observations as the carrier concentrations can be precisely controlled in high-quality samples.

The key findings in our Hall measurements are summarized in Fig. 3, where we plot the gate dependent $R_{xx}$ and the corresponding Hall coefficient $R_H$ measured in device D1 and D2 under decreasing $T$. Determining $R_H$ of an insulator is challenging as a slight misalignment of the Hall probes will result in strong mixing signals from $R_{xx}$. We apply the standard anti-symmetrization process to extract $R_H$ by sweeping the out-of-plane field ($B$) in both directions, namely, $R_H \equiv \alpha\, dR_{xy}^{as}/dB$, where $R_{xy}^{as}$ is the asymmetric component of $R_{xy}$ (supplementary section 9) and $\alpha$ is a factor accounting for the device geometry (supplementary section 3 & 4). At high $T$ (~ 200 K), the monolayer behaves like a semimetal where a sign change of $R_H$ occurs in the hole-dominant side away from the CNP (Fig. 3b and d), agreeing with the semiclassical expectation that $R_H$ changes its sign at $n_h = (u_e/u_h)^2\, n_e$. Here $n_h$ ($n_e$) is the density of holes (electrons) and $u_h$ ($u_e$) is the corresponding mobility. The anomaly is that while $R_{xx}$ rapidly increases with either lowering $T$ or reducing doping to the CNP, $R_H$ is however quite insensitive to the insulating $R_{xx}$ down to the lowest $T$ for reliable Hall measurements. For instance, while the sharp $R_{xx}$ peak reaches a value of ≈ 0.6 MΩ at 80 K in D1, the overall shape and values of $R_H$ do not deviate much from its semi-metallic profile at high $T$, despite being much noisier. The behavior is in sharp contrast to the $R_H$ of a band insulator, which is expected to develop a rapid sign change accompanied with the $R_{xx}$ peak and a diverging behavior when approaching it from either side (see green curves in Fig. 3b and d and the observations in gapped graphene[43,44]). While the observed $R_H$ approaches the green curve in the doped metallic regime, the Hall characteristics of a conventional insulator are clearly absent in our devices. The two devices with different disorder strength show consistent behaviors. While disorders have produced a dramatic change on $R_{xx}$, including an order of magnitude change in the peak width, their effect on $R_{xy}$ is nevertheless qualitatively insignificant. If one only looks at the $R_{xy}$ data (Fig. 3 b and d), there is no sign of the formation of an insulator state. By contrast, the $R_{xx}$ data has clearly shown an insulator state at a specific doping (the CNP). This Hall response of the monolayer insulator behaves like a semimetal, rather than a conventional insulator. These observations of the decoupled behaviors in $R_{xx}$ and $R_{xy}$ suggest that the appearance of the low temperature insulator state is not due to the depletion of carriers; instead, it is due to the correlations between electrons and holes at CNP (i.e., excitonic pairing between an exactly equal number of electrons and holes).

We further confirm the correlated nature of the insulator by gate-tuned tunneling spectroscopy based on both vdW tunneling spectroscopy and STM. In vdW tunneling devices, we fabricate narrow graphite fingers (≈ 100 nm wide) underneath the $WTe_2$ flake separated by a few-layer hBN tunneling barrier (Fig. 4a and Fig. S4). At large bias, our devices show consistent results with earlier STM studies on samples grown on graphene[27,34] (supplementary section 10). The goal here is to reliably measure the low energy behavior (< 100 meV) in the pristine monolayer, which is key to understanding the insulating state. Fig. 4b-d plot differential conductance $dI/dV$ under varying dc sample bias $V_b$ and the top gate voltage $V_{tg}$, measured in a vdW device (D4) at selected $T$. While almost no feature appears in the map at high $T$ (except the gate independent phonon characteristics[45], see supplementary section 10), a tunneling gap centered at zero bias clearly develops at low $T$ near CNP ($V_{tg} \approx 0$ V). The gap, with a U-shape and a size of about 47 meV (supplementary section 11), closes when the monolayer is doped with either electrons or holes (Fig. 4d), confirming the metal-insulator transition observed in transport.

Qualitatively similar results are observed in our STM measurements. For the STM device (D7), monolayer hBN was used to cover the $WTe_2$, protecting the flake but still allowing electrons to tunnel through (Fig. 4e and supplementary section 2). Fig. 4f and g present the corresponding STM $dI/dV$ map and typical spectra taken at selected gate voltages ($T = 1.4$ K). At high doping, the spectra feature a V-shaped linear suppression of the $dI/dV$ signal towards Fermi energy, consistent with the presence of a Coulomb gap[37] in the metallic regime due to finite disorders in the sample. As the doping is reduced towards the CNP, the linear suppression consolidates, i.e., $dI/dV$ at zero bias reaches zero, and then transforms into a fully depleted U-shaped hard gap at the CNP with a size of about 91 meV, again signifying the metal-insulator transition (Fig. 4g). Consistent results are observed when the STM tip is engaged at different sample locations (supplementary section 12), as well as in two additional vdW tunneling devices (D5 & D6, supplementary section 13). Our data are distinct from the previous STM study[34] on the $WTe_2$/graphene stacks with doping controlled by atomic dopants, where a different screening environment and strong disorders are expected. There, a V-shaped soft gap was observed at all doping and no clear signature of the metal-insulator transition was seen[34]. We note that the graphene substrate is known to significantly reduce exciton binding energy, as demonstrated for optical excitons in 2D semiconductors[46]. Our tunneling experiments on pristine monolayer $WTe_2$ demonstrate a clear contradiction to any band insulator scenario, including a strain-induced band insulator[47], as a conventional band gap, with or without a correlation-modified gap size, will shift its bias position when the Fermi level is tuned away from the gap. Our data reveals a correlation-induced gap that is always pinned at zero bias. This observation, together with the gate-induced metal-insulator transition, is consistent with the presence of an intrinsic EI gap at the CNP (see more discussion in supplementary section 14).

We summarize our observations by sketching a low-$T$ electronic phase diagram of the monolayer under varying $D$ and $n_g$ (Fig. 4h). Metallic phases reside at high doping in either electron- or hole-dominant side, while reducing $n_g$ leads to insulating behaviors. Our systematic transport and tunneling experiments on samples of controlled quality imply an intrinsic correlated insulator phase at the CNP in the clean limit of the pristine WTe$_2$ monolayer. The results rule out the scenario of a band insulator and support the presence of an EI phase. Electron localization fails to account for the observations at the CNP, yet it likely plays a role in real samples and may coexist with the excitonic order in the intermediate regime of the diagram. Our results call for future efforts in further understanding the nature of the ground states in monolayer WTe$_2$, where the interplay between the EI phase and topology, as well as superconductivity, may open a new avenue for exploring correlated quantum states. The observations also identify WTe$_2$ monolayer as a promising material platform for constructing quantum devices utilizing coherent excitons in the ground state.

**Acknowledgements**

We acknowledge helpful discussions with N. P. Ong and P. A. Lee. Work in Wu lab was primarily supported by NSF through a CAREER award to S. W. (DMR-1942942). Device fabrication was supported by NSF-MRSEC through the Princeton Center for Complex Materials NSF-DMR-1420541 & DMR-2011750. S.W. and L.M.S. acknowledge the support from Eric and Wendy Schmidt Transformative Technology Fund at Princeton. Part of the measurements was performed at the National High Magnetic Field Laboratory, which is supported by NSF Cooperative Agreement No. DMR-1644779 and the State of Florida. Work in Yazdani lab was primarily supported by the Gordon and Betty Moore Foundation's EPiQS initiative grants GBMF4530, GBMF9469, and DOE-BES grant DE-FG02-07ER46419. Other support for the experimental work by A. Y. was provided by NSF-DMR-1904442, ExxonMobil through the Andlinger Center for Energy and the Environment at Princeton, and the Princeton Catalysis Initiative. B. A. B is supported by the Department of Energy Grant No. DE-SC0016239, the Schmidt Fund for Innovative Research, Simons Investigator Grant No. 404513, the Packard Foundation for the numerical work. The analytical part was supported by the National Science Foundation EAGER Grant No. DMR- 1643312, BSF Israel US foundation No. 2018226, ONR No. N00014-20-1-2303, and the Princeton Global Network Funds. Additional support to B. A. B was provided by the Gordon and Betty Moore Foundation through Grant GBMF8685 towards the Princeton theory program. B.J. acknowledges funding through a postdoctoral fellowship of the Alexander-von-Humboldt foundation. K.W. and T.T. acknowledge support from the Elemental Strategy Initiative conducted by the MEXT, Japan, Grant Number JPMXP0112101001, JSPS KAKENHI Grant Number JP20H00354 and the CREST(JPMJCR15F3), JST. F.A.C. and R.J.C. acknowledge support from the ARO MURI on Topological Insulators (grant W911NF1210461). S.L, S.K., and L.M.S. acknowledge support from the Gordon and Betty Moore Foundation through Grant GBMF9064 awarded to L.M.S.

**Author Contributions**

S.W. supervised transport and vdW tunneling studies. A. Y. supervised STM studies. P.W. and G.Y. fabricated transport devices. Y.J. fabricated the vdW tunneling devices, assisted by P.W., G.Y., M.O., N.F., and B.J. Y.J., P.W., and S. W. performed transport and vdW tunneling measurements and analyzed data. C-L.C., Y.J., P.W. and X.L. fabricated the STM device. C-L.C., G.F., X.L., and B.J. performed STM measurements and analyzed data. Z. S., F. X., Y.X. and B. A.

B provided theoretical support. S.L., S.K., L.M.S., F.A.C. and R.J.C. grew and characterized bulk WTe$_2$ crystals. K.W. and T.T. provided hBN crystals. All authors discussed the result and contributed to the writing of the paper.

**Competing Interests**

The authors declare no competing financial interests.

**Figure Captions**

**Figure 1 | Possible scenarios of the ground states at CNP in monolayer WTe$_2$. a,** SOC-induced band insulator. The sketch depicts the low energy bands with and without SOC. The dashed black line indicates the Fermi level. **b,** A semimetal phase is formed if the SOC cannot gap out the entire Fermi surface, resulting in one hole-pocket and two electron-pockets in the system. **c,** Electron localization can produce an insulating phase from a 2D semimetal via a disorder-induced Coulomb gap. **d,** An intrinsic excitonic insulator phase that hosts mobile excitons in the ground state.

**Figure 2 | The insulating state at charge neutrality in monolayer WTe$_2$. a,** Cartoon illustration of the transport device, where the electrodes contact the WTe$_2$ monolayer bulk without touching its edges. **b,** An optical image of a typical device (D1). **c,** Four-probe resistance mapped under varying top ($V_{tg}$) and bottom ($V_{bg}$) gates, taken from D1 at 70 K. The measurement configuration is shown in **d**. **d,** $R_{xx}$ as a function of gate-induced density $n_g$ (defined in the main text) taken from three devices (D1-D3) under the same measurement configuration shown as the right inset. The $n_g$ traces are selected to cross the most insulating regime in their respective dual-gate resistance maps. The displacement field is fixed for D1 and D3 curves, while $V_{bg}$ is set at 0 V for the D2 curve. The choice was made to maximize the scan range of $n_g$ for each device. Qualitatively similar curve for D2 is obtained if the displacement field is fixed instead. Left inset displays the temperature effect on the $R_{xx}$ curves of D1. **e,** $R_{xx}$ as a function of $T$ taken from D1 and D2 at the most insulating region in their respective resistance maps (at CNP). **f,** Displacement field effect on the $R_{xx}$ curve, taken from D1 along the charge neutrality line in **c**. The value of $D$ for each curve is indicated next to it.

**Figure 3 | Hall anomaly in the monolayer insulator. a,** $R_{xx}$ as a function of $n_g$ with fixed displacement field, taken from D1 at various $T$ indicated. Inset shows the measurement configuration. **b,** The corresponding Hall response, $dR_{xy}^{as}/dB$, of the same device measured down to 80 K, below which the Hall measurement around the insulating state is no longer reliable, indicated by the enhanced noise fluctuations at low $T$. **c** and **d**, $R_{xx}$ and $dR_{xy}^{as}/dB$ measured in D2 where $n_g$ is varied by tuning $V_{tg}$ with a fixed $V_{bg}$ = 0 V. The Hall coefficient $R_H$ (right axis in **b** and **d**) is calibrated by a device-specific geometry factor (supplementary section 3 & 4). The green lines in **c** and **d** indicate the expected $R_H$ of a band insulator, i.e., -1/$e(n_g-n_{g0})$, where $n_{g0}$ is the location of CNP labeled by the dashed vertical line.

**Figure 4 | Signature of correlations and the metal-insulator transition revealed by tunneling spectroscopy. a,** Schematic for the vdW tunneling devices used in our measurements, where a few-layer hBN flake serves as the tunneling barrier. Top and bottom graphite gates are employed. The measurement configuration is sketched as well. **b-d**, d$I$/d$V$ tunneling spectra as a function of

sample bias $V_b$ and top gate voltage $V_{tg}$, taken from device D4 at selected $T$. The bottom gate $V_{bg}$ is fixed at 3 V to enhance the conductivity of the monolayer outside the tunneling junction. The tilted lines appeared in the insulating regime may arise due to the effect of probe-induced doping. **e**, Schematic for the STM device (D7) used in our study, where a monolayer hBN is used to protect the WTe$_2$ flake while allowing tunneling experiments. A silicon bottom gate is employed. **f**, Gate-tuned tunneling spectra taken from D7 in a 1.4 K STM system. **g**, d$I$/d$V$ spectra at selected $V_g$ indicated by the color arrows in **f**. The curves are shifted vertically to enhance visibility, where dashed lines indicate the corresponding zero conductance respectively for each curve. **h**, A sketch of the electronic phase diagram of the monolayer at low $T$ under varying $D$ and $n_g$. At ultralow temperatures, the metallic phase at the electron side will turn into a superconductor (SC).

**Methods**

**Crystal Growth and Sample Fabrication**

The vapor-grown WTe$_2$ bulk crystals were grown using methods that were described in earlier works[36]. The flux-grown crystals were synthesized by a solid-state reaction using Te as the flux. The starting material Te (99.9999%, Alfa Aesar) was firstly purified to remove oxygen contaminations and then mixed with W (99.9%, Sigma-Aldrich) in a molar ratio of 98.8:1.2. The raw materials mixture was sealed in an evacuated quartz ampoule and heated to 1020°C over a period of 16 hours and slowly cooled down to 700 °C at a rate of 1.2°C/hour and then to 540°C at a rate of 2°C/hour. The crystals were obtained by a decanting procedure in a centrifuge. The detailed 2D device fabrication procedures and parameters are described in supplementary section 2.

**Transport and vdW tunneling Measurement**

The transport and vdW tunneling measurements were performed in a Quantum Design Dynacool system, a Quantum Design Opticool, or in a dilution refrigerator (Blufors LD400). The dilution refrigerator was equipped with low-pass RC filters at room temperature and low-pass LC filters (Mini-Circuits VLFX-80+) at the mixing chamber stage. The measurement was conducted with standard lock-in technique at a low frequency (2 – 15 Hz). The ac excitation current used in transport measurement was on the order of 1 nA.

**STM Measurement**

After swiftly mounting the chip onto the sample holder, the sample was transferred into an ultra-high vacuum chamber (< 10$^{-10}$ torr) and baked at 350 °C overnight to remove residues on the surface of the vdW stack. The device was then installed into a home-built 1.4 K STM and the measurement was performed with a tungsten tip prepared on Cu (111) surface. The differential conductance was acquired by using the standard lock-in techniques with an ac excitation voltage at a frequency of 4000 Hz.

**Data Availability**

The data that support the plots within this paper are available at https://doi.org/10.7910/DVN/FFGQOX. Other data that support the findings of this study are available from the corresponding author upon reasonable request.

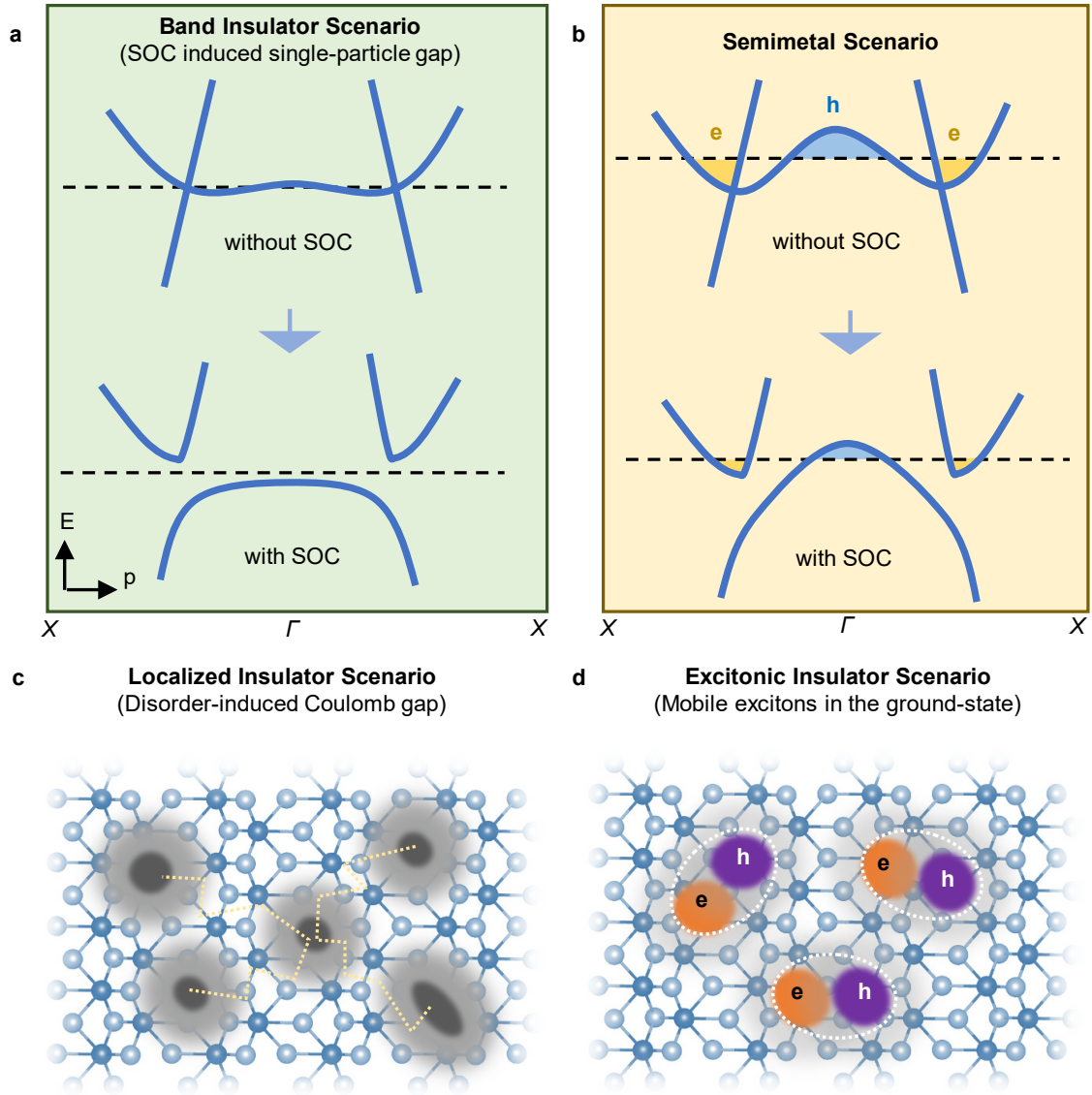

**Figure 1 | Possible scenarios of the ground states at CNP in monolayer WTe$_2$. a,** SOC-induced band insulator. The sketch depicts the low energy bands with and without SOC. The dashed black line indicates the Fermi level. **b,** A semimetal phase is formed if the SOC cannot gap out the entire Fermi surface, resulting in one hole-pocket and two electron-pockets in the system. **c,** Electron localization can produce an insulating phase from a 2D semimetal via a disorder-induced Coulomb gap. **d,** An intrinsic excitonic insulator phase that hosts mobile excitons in the ground state.

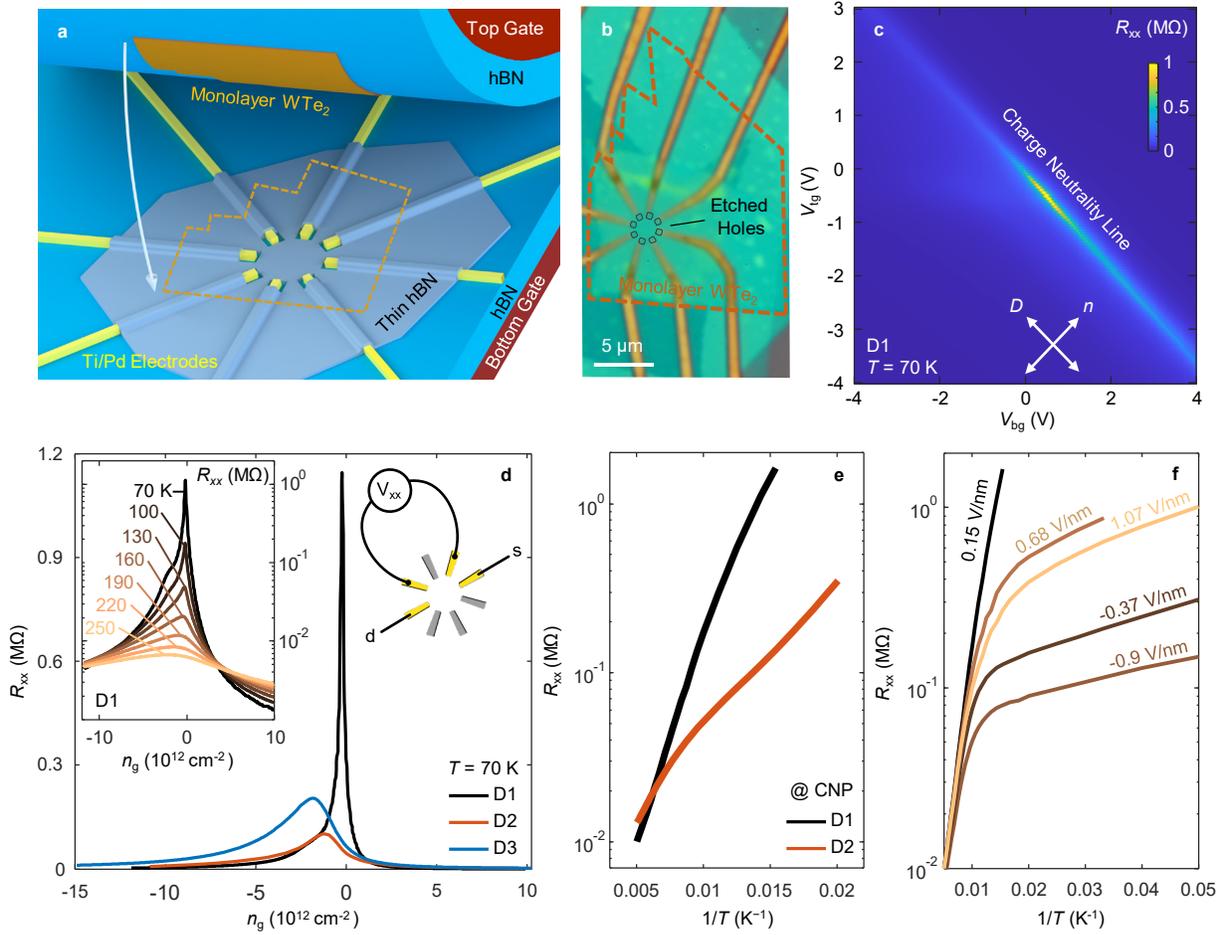

**Figure 2 | The insulating state at charge neutrality in monolayer WTe$_2$. a,** Cartoon illustration of the transport device, where the electrodes contact the WTe$_2$ monolayer bulk without touching its edges. **b,** An optical image of a typical device (D1). **c,** Four-probe resistance mapped under varying top ($V_{tg}$) and bottom ($V_{bg}$) gates, taken from D1 at 70 K. The measurement configuration is shown in **d**. **d,** $R_{xx}$ as a function of gate-induced density $n_g$ (defined in the main text) taken from three devices (D1-D3) under the same measurement configuration shown as the right inset. The $n_g$ traces are selected to cross the most insulating regime in their respective dual-gate resistance maps. The displacement field is fixed for D1 and D3 curves, while $V_{bg}$ is set at 0 V for the D2 curve. The choice was made to maximize the scan range of $n_g$ for each device. Qualitatively similar curve for D2 is obtained if the displacement field is fixed instead. Left inset displays the temperature effect on the $R_{xx}$ curves of D1. **e,** $R_{xx}$ as a function of $T$ taken from D1 and D2 at the most insulating region in their respective resistance maps (at CNP). **f,** Displacement field effect on the $R_{xx}$ curve, taken from D1 along the charge neutrality line in **c**. The value of $D$ for each curve is indicated next to it.

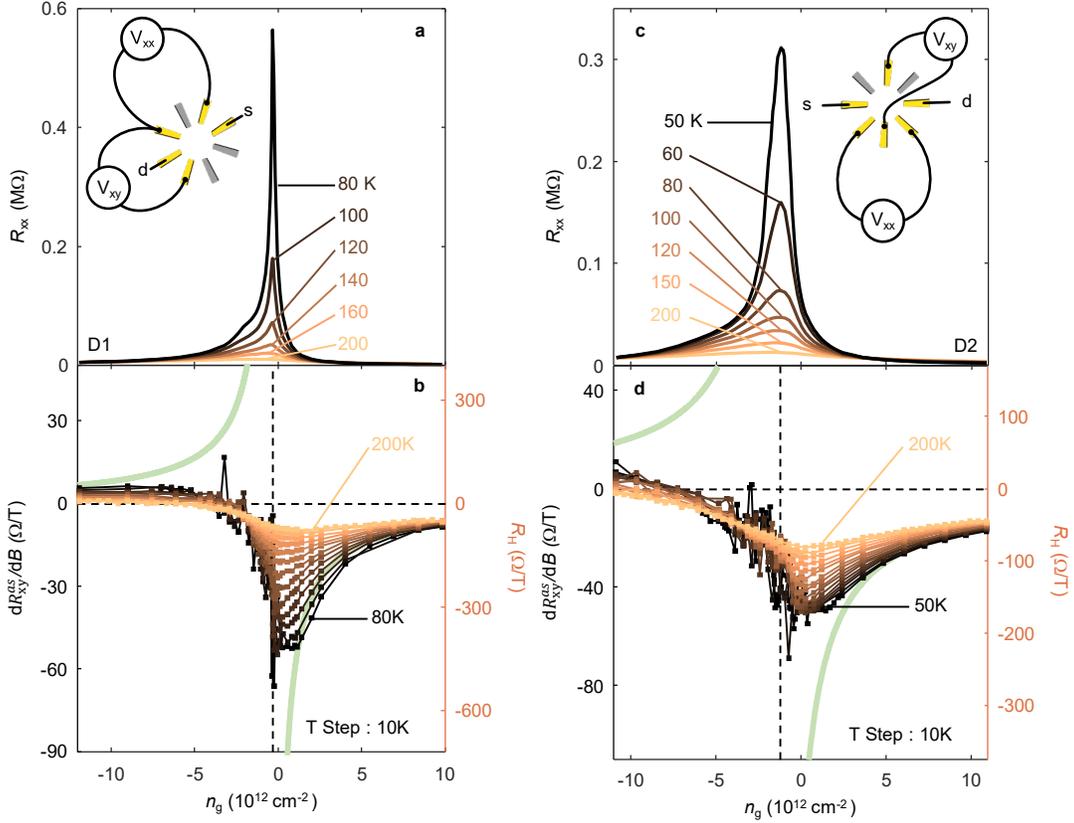

**Figure 3 | Hall anomaly in the monolayer insulator. a,** $R_{xx}$ as a function of $n_g$ with fixed displacement field, taken from D1 at various $T$ indicated. Inset shows the measurement configuration. **b**, The corresponding Hall response, $dR_{xy}^{as}/dB$, of the same device measured down to 80 K, below which the Hall measurement around the insulating state is no longer reliable, indicated by the enhanced noise fluctuations at low $T$. **c** and **d**, $R_{xx}$ and $dR_{xy}^{as}/dB$ measured in D2 where $n_g$ is varied by tuning $V_{tg}$ with a fixed $V_{bg} = 0$ V. The Hall coefficient $R_H$ (right axis in **b** and **d**) is calibrated by a device-specific geometry factor (supplementary section 3 & 4). The green lines in **c** and **d** indicate the expected $R_H$ of a band insulator, i.e., $-1/e(n_g-n_{g0})$, where $n_{g0}$ is the location of CNP labeled by the dashed vertical line.

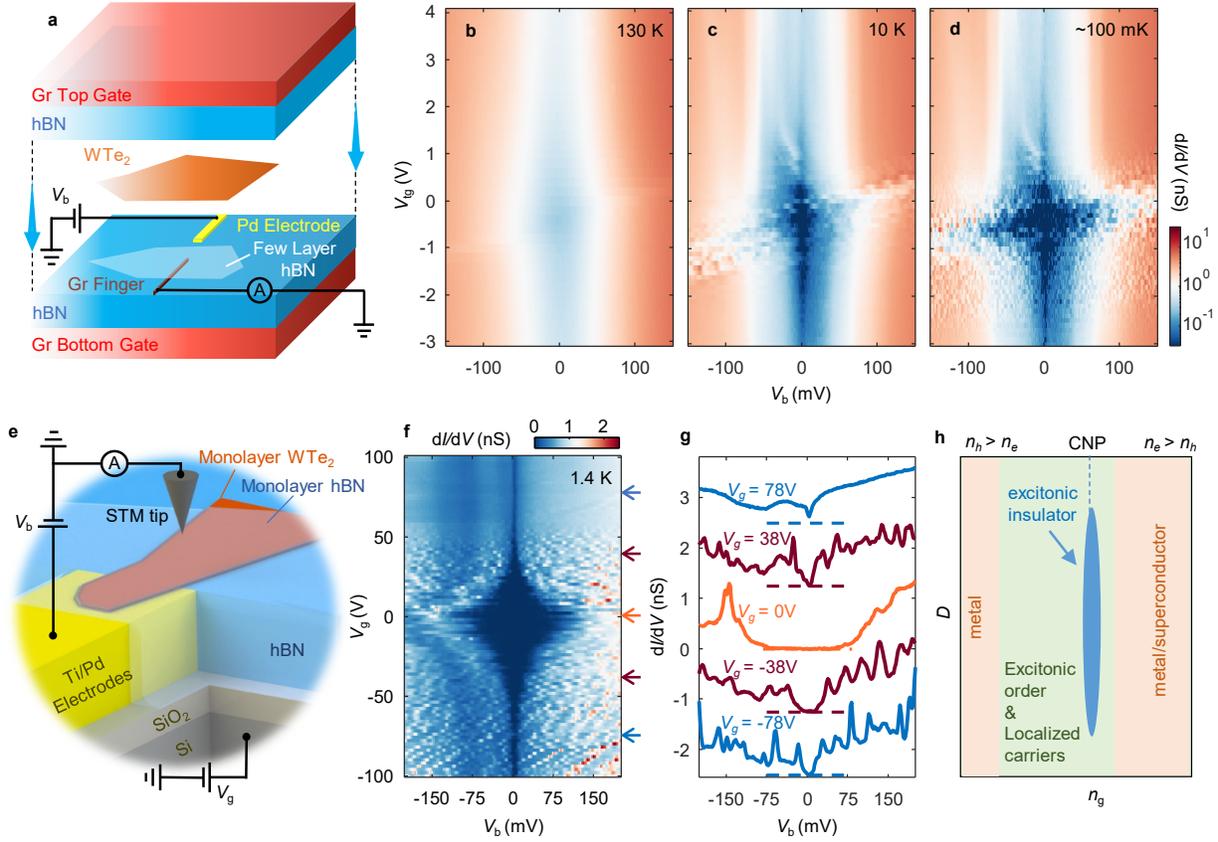

**Figure 4 | Signature of correlations and the metal-insulator transition revealed by tunneling spectroscopy. a,** Schematic for the vdW tunneling devices used in our measurements, where a few-layer hBN flake serves as the tunneling barrier. Top and bottom graphite gates are employed. The measurement configuration is sketched as well. **b-d,** d$I$/d$V$ tunneling spectra as a function of sample bias $V_b$ and top gate voltage $V_{tg}$, taken from device D4 at selected $T$. The bottom gate $V_{bg}$ is fixed at 3 V to enhance the conductivity of the monolayer outside the tunneling junction. The tilted lines appeared in the insulating regime may arise due to the effect of probe-induced doping. **e,** Schematic for the STM device (D7) used in our study, where a monolayer hBN is used to protect the WTe$_2$ flake while allowing tunneling experiments. A silicon bottom gate is employed. **f,** Gate-tuned tunneling spectra taken from D7 in a 1.4 K STM system. **g,** d$I$/d$V$ spectra at selected $V_g$ indicated by the color arrows in **f**. The curves are shifted vertically to enhance visibility, where dashed lines indicate the corresponding zero conductance respectively for each curve. **h,** A sketch of the electronic phase diagram of the monolayer at low $T$ under varying $D$ and $n_g$. At ultralow temperatures, the metallic phase at the electron side will turn into a superconductor (SC).

# Evidence for a Monolayer Excitonic Insulator

Yanyu Jia[1,#], Pengjie Wang[1,#], Cheng-Li Chiu[1,#], Zhida Song[1], Guo Yu[1,2], Berthold Jäck[1], Shiming Lei[3], Sebastian Klemenz[3], F. Alexandre Cevallos[3], Michael Onyszczak[1], Nadezhda Fishchenko[1], Xiaomeng Liu[1], Gelareh Farahi[1], Fang Xie[1], Yuanfeng Xu[4], Kenji Watanabe[5], Takashi Taniguchi[6], B. Andrei Bernevig[1], Robert J. Cava[3], Leslie M. Schoop[3], Ali Yazdani[1,*], Sanfeng Wu[1,*]

[1] Department of Physics, Princeton University, Princeton, New Jersey 08544, USA
[2] Department of Electrical Engineering, Princeton University, Princeton, New Jersey 08544, USA
[3] Department of Chemistry, Princeton University, Princeton, New Jersey 08544, USA
[4] Max Planck Institute of Microstructure Physics, 06120 Halle, Germany
[5] Research Center for Functional Materials,
National Institute for Materials Science, 1-1 Namiki, Tsukuba 305-0044, Japan
[6] International Center for Materials Nanoarchitectonics,
National Institute for Materials Science, 1-1 Namiki, Tsukuba 305-0044, Japan
[#]These authors contributed equally to this work.
[*] Email: sanfengw@princeton.edu; yazdani@princeton.edu

**Contents**

# 1. Theoretical Modeling of the excitonic insulator phase in WTe$_2$ monolayer

The excitonic pairing in monolayer WTe$_2$ involves two electron pockets and one hole pocket, which is nontrivially different from the simple two-pocket (one of electrons, one of holes) system. The topological band inversion in the system has also a unique character compared to the regular EI problem. To theoretically model the EI phase in monolayer WTe$_2$, we apply self-consistent Hartree-Fock mean field calculations with changing interaction strength. We find that weak interaction favors the EI phase with a small gap of order 100 meV, whereas strong interaction favors a trivial insulator with large gap of order 1 eV, which is equivalent to an atomic insulator. A first order phase transition happens at the critical interaction 1.25 eV. (Interaction strength is defined below). By computing the inversion eigenvalues of the EI phase, we confirm that the EI phase is a topological insulator, with a gap size affected by both the excitonic order parameter and the strength of spin-orbit coupling (SOC). The interaction strength in the experiment is estimated as 0.925 eV and hence stabilizes the EI phase.

We use the following kp model for the band structure

$$H_0 = \left(a\mathbf{k}^2 + b\mathbf{k}^4 + \frac{\delta}{2}\right)\begin{pmatrix}1 & 0 & 0 & 0\\ 0 & 1 & 0 & 0\\ 0 & 0 & 0 & 0\\ 0 & 0 & 0 & 0\end{pmatrix} + \left(-\frac{\mathbf{k}^2}{2m} - \frac{\delta}{2}\right)\begin{pmatrix}0 & 0 & 0 & 0\\ 0 & 0 & 0 & 0\\ 0 & 0 & 1 & 0\\ 0 & 0 & 0 & 1\end{pmatrix} + v_x k_x \tau_x s_y + v_y k_y \tau_y s_0,$$

where $\tau_x, \tau_y, s_y$ are Pauli-matrices and $s_0$ is two-by-two identity matrix. The basis set is $|d\uparrow\rangle$, $|d\downarrow\rangle$, $|p\uparrow\rangle$, $|p\downarrow\rangle$, where $p/d$ represent the atomic orbital and $\uparrow/\downarrow$ represent the spin. This model has inversion symmetry ($P = \tau_z$) and time-reversal symmetry ($T = is_y K$). We take the parameters

$$a = -3,\ b = 18,\ m = 0.03,\ \delta = -0.9,\ v_x = 0.5,\ v_y = 3.$$

The energy is in units of eV. The unit cell, Brillouin zone (BZ) are shown in Fig. S1a. The wavevector connecting the hole pocket and the electron pocket is $q_c$, which is approximately 1/3 of the line ΓX. Thus the excitonic order parameter should the have the wavevector $q_c$ and break the translation symmetry in the x-direction.

Before we move to the interacting Hamiltonian, we first calculate the bare susceptibility using the kp model. The susceptibility is given by the following formula:

$$\Pi(\mathbf{q}) = -\frac{T}{N}\sum_{\omega_n,\mathbf{k}} \frac{1}{-i\omega_n + h(\mathbf{k})}\frac{1}{-i\omega_n + h(\mathbf{k}+\mathbf{q})},$$

in which $\omega_n$ is fermionic Matsubara frequencies, $T$ is the temperature, and $N$ is number of momenta in the Brillouin zone. We then numerically evaluated the trace of the susceptibility, as shown in Fig. S1b. It can be noticed that the susceptibility develops peaks at around $q = \pm q_c$, which suggests a possible excitonic density wave order parameter.

The interaction Hamiltonian has the form

$$H_{\text{int}} = \frac{1}{2\Omega N}\sum_{kpq}\sum_{\alpha,\beta} V(q) c^+_{k+q,\alpha} c^+_{p-q,\beta} c_{p,\beta} c_{k,\alpha}.$$

where $N$ is the number of momenta considered, $k, p, q$ index momenta in the entire BZ, $\Omega$ is the area of the unit cell, and $\alpha, \beta$ are the orbital/spin indices. $V(q)$ is the Fourier transformation of the double-gate-screened Coulomb interaction

$$V(q) = \pi \xi^2 V_\xi \frac{\tanh \xi q/2}{\xi q/2}, \qquad V_\xi = \frac{e^2}{4\pi\epsilon\xi}.$$

where $\xi \approx 25$nm is the distance between the two gates used in the experiments, $\epsilon = 3.5$ is the dielectric constant of hNB. Since we use a kp model for calculation, we need to set a momentum cutoff and only consider momenta within the cutoff. Thus, we rewrite the interaction as

$$H_{\text{int}} = \frac{1}{2\Omega' N_{kp}} \sum_{kpq \in A_{kp}} \sum_{\alpha,\beta} V(q) c^+_{k+q,\alpha} c^+_{p-q,\beta} c_{p,\beta} c_{k,\alpha}, \qquad \Omega' = \Omega \frac{A_{BZ}}{A_{kp}},$$

with $N_{kp}$ being the number of momenta within the cutoff, $A_{BZ}$ the area of the BZ, $A_{kp}$ the area of the region within the cutoff. Since $N_{kp}/A_{kp}=N/A_{BZ}$, we keep the same density of the momenta in the momentum cutoff region as that in the full Brillouin zone. We further rewrite the interaction as

$$H_{\text{int}} = \frac{1}{2N_{kp}} \sum_{kpq \in A_{kp}} \sum_{\alpha,\beta} U(q) c^+_{k+q,\alpha} c^+_{p-q,\beta} c_{p,\beta} c_{k,\alpha}$$

$$U(q) = U_0 \frac{\tanh \xi q/2}{\xi q/2}, \qquad U_0 = \frac{\pi \xi^2 V_\xi}{\Omega'}.$$

$U_0$ has the dimension of energy. In the following, we fix $\xi = 25$nm and use $U_0$ as a tuning parameter. We refer to $U(q_c)$, i.e., the interaction at the order wave-vector, as the interaction strength. The momentum cutoffs for the Hartree-Fock calculation are $|k_x| \leq \frac{3}{2}|q_c|$, $|k_y| \leq |k_{Y'}|$, where $Y'$ is the midpoint between $\Gamma$ and $Y$. Since the order parameter will have the wavevector $q_c$, we fold the BZ in the x-direction, i.e., identify $k$ and $k + q_c$ as the same momentum. The folded BZ is shown in Fig. S1a. The unfolded and folded band structure without interaction ($U = 0$) is shown in Fig. S1c and d, along with the inversion eigenvalues.

In Fig. S1e and f, we plot the band structures of the exciton ordered phase ($U(q_c) = 1.125$ eV) and the non-exciton phase ($U(q_c) = 2.25$ eV). In the ordered phase, the hole pocket and electron pocket hybridize and a gap of order 100 meV is opened. Since the band inversion between $d$ and $p$ orbitals is preserved, the ordered phase is in fact a topological insulator and the gap size is mainly determined by the spin-orbit coupling strength, which is ~150 meV in the kp model. In the non-exciton phase, a huge gap (~ 2 eV) is opened and the band inversion is reverted. Thus, the non-exciton phase is a trivial insulator.

In Fig. S2a, the Hartree-Fock total energy is plotted as a function of the interaction strength. We observe a sharp change of the slop of the energy around $U_c(q_c) = 1.25$eV, implying a first order phase transition. To identify the two phases, we compute the averaged order parameter

$$\Delta = \sqrt{\frac{1}{N_{kp}} \sum_{k \in A_{kp}} \sum_{\alpha\beta} |\langle c^+_{k,\alpha} c_{k+q_c,\beta} \rangle|^2}.$$

As shown in Fig. S2b, $\Delta$ jumps to zero at $U_c(q_c) = 1.25$ eV. Thus $U(q_c) < U_c(q_c)$ corresponds to the ordered phase. With the lattice constants given in Fig. S1a, the cutoff defined above, and $\xi = 25$ nm, $\epsilon = 3.5$, the interaction strength in the real device is estimated as $U_{\exp}(q_c) = 0.925$ eV, which confirms the experimental observation of the EI phase.

We further comment that the EI phase here is also a TI and the indirect gap of the TI is bounded by the SOC. It (approximately) equals the SOC if the indirect gap is the same as the smallest direct gap. If the system does not have SOC, then the equivalent time-reversal symmetry in each spin sector is the complex conjugation, which squares to 1. Such a time-reversal symmetry does not protect topological phases. One can define a $Z_2$ index

$$(-1)^\delta = \prod_K \prod_n \xi_{K,n}$$

where K indexes the four inversion-invariant momenta in the 2D Brillouin zone, n indexes the occupied bands in the spin-up sector, and $\xi_{K,n}$ is the inversion eigenvalue of the n-th occupied state at K. $\delta=1$ corresponds to a Dirac semimetal phase. When SOC presents, the time-reversal symmetry is $is_y K$, which squares to -1 and protects the 2D topological insulator. One can define the same index $\delta$, but now $\delta=1$ corresponds to the topological insulator instead of the Dirac semimetal, provided that the bands have an indirect gap. If a system with $\delta=1$ does not have an indirect gap, we could still think the system as a topological insulator if we occupied the states below the direct gap at each momentum.

We consider a model with $\delta=1$ and a tunable SOC strength. When SOC is zero, the model is a Dirac semimetal. As we continuously increase SOC, the model becomes a topological insulator (at least in the sense of the direct gap). Thus, a direct gap must be opened by SOC and the direct gap is of the order of SOC. The indirect gap is in general smaller than the direct gap and could even be zero (e.g., Fig. S1c) in some cases. Thus, the SOC is an upper bound of the TI. In Fig. S1e we see that the indirect gap equals the direct gap, which has the same order as SOC.

We comment that our calculation implies the presence of a density wave order in the insulating state and hence call for further efforts to search for experimental evidence of the order. We also studied the role of displacement field ($D$) on the one-body band structure through first-principle calculations. In Fig. S3a and b, the band structures without folding the BZ with $D = 0$ and $D = 0.5$ V/nm are plotted respectively. $D$ breaks the inversion symmetry $P$ and hence the combination of inversion and time-reversal $PT$. We find that a finite $D$ splits the spin degeneracy of the bands and also gives rise to larger Fermi pockets for both electrons and holes simultaneously at charge neutrality.

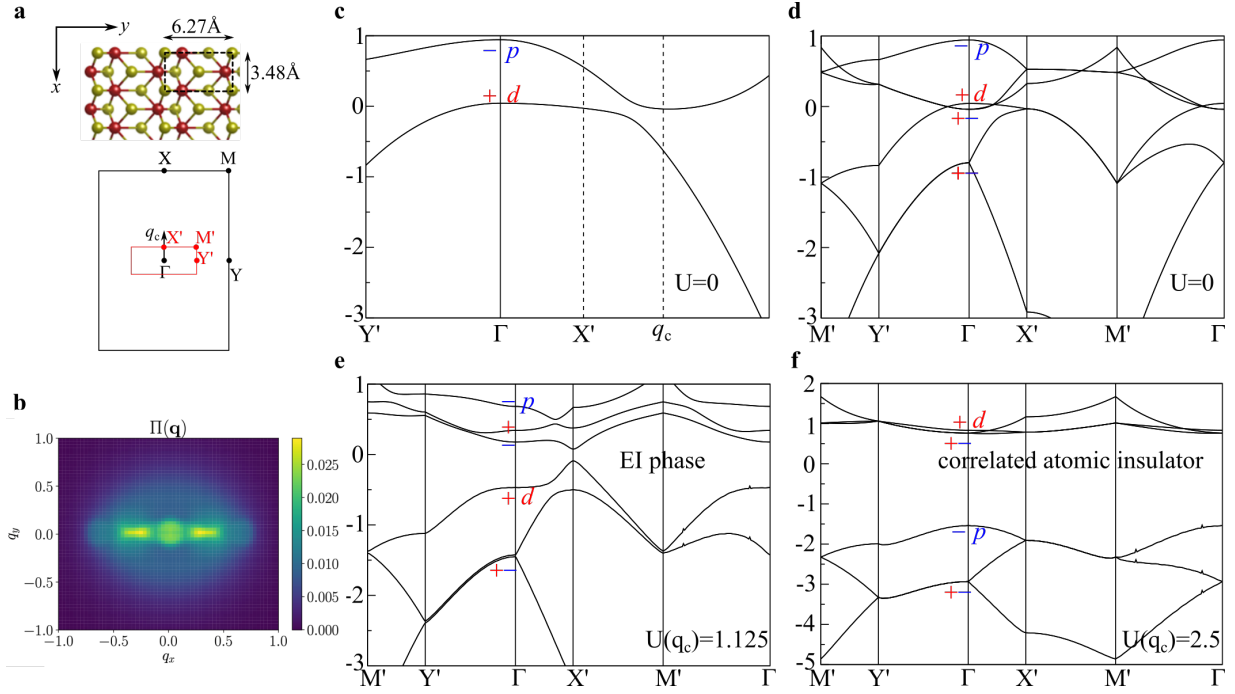

**Fig. S1. The band structure of EI phase. a,** Top: the lattice and the lattice constants. Bottom: the Brillouin zone and the folded (red) Brillouin zone. $q_c$ point is the minimum of the electron band. X' is half of $q_c$. Y' is the midpoint between G and Y. **b,** The real part of maximal eigenvalue of the susceptibility of the kp model at each momentum. We note that peaks of the of susceptibility develop at $q = \pm q_c$, which suggests an excitonic density wave-vector. **c,** The band structure of the kp model. **d,** The folded band structure of the kp model. **e,** The band structure of EI phase, with $U(q_c) = 1.125$eV. **f,** The band structure of the correlated atomic insulator phase, with $U(q_c) = 2.25$eV. The "+/-" signs are the inversion eigenvalues of the corresponding Bloch states. The states labeled with "p/d" are mainly contributed by the p and d orbitals, respectively. The energies are in units of eV. The first-principle calculations are performed on the Vienna ab-initio simulation package (VASP)[1,2] and the generalized gradient approximation (GGA)[3] with the Perdew-BurkeErnzerhof (PBE)[3,4] type exchange-correlation potential is adopted. We use the Brillouin zone (BZ) sampling with a 11×11×1 mesh in the self-consistent calculations. The plane-wave cutoff energy is 300 eV. SOC is included in the calculations.

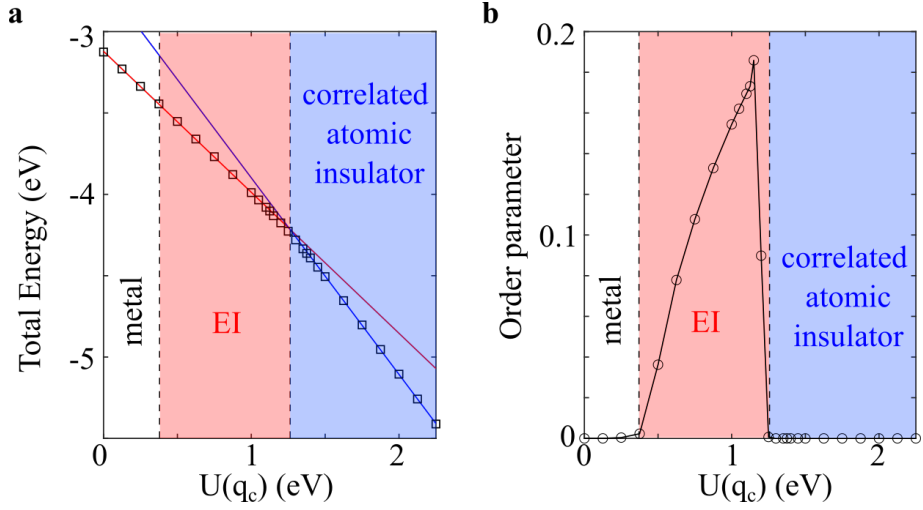

**Fig. S2. The total energy and the order parameter as functions of interaction strength $U_0$.** The EI phase is stabilized when $U_{c1}(q_c)<U(q_c)<U_{c2}(q_c)$, and the correlated atomic insulator phase is stabilized when $U(q_c) > U_{c2}(q_c)$, where $U_{c1}(q_c)=0.35$eV and $U_{c2}(q_c)=1.25$eV. **a,** There is a sharp change of the slop of the energy at $U_{c2}(q_c)$, implying a first order phase transition. The red and blue lines are the linear fittings of the energies in EI phase and correlated atomic insulator phase, respectively. **b,** The order parameter becomes nonzero at $U_{c1}(q_c)$ and jumps to zero at $U_{c2}(q_c)$. The experimental value of $U(q_c)$ is estimated to be ~ 0.925 eV.

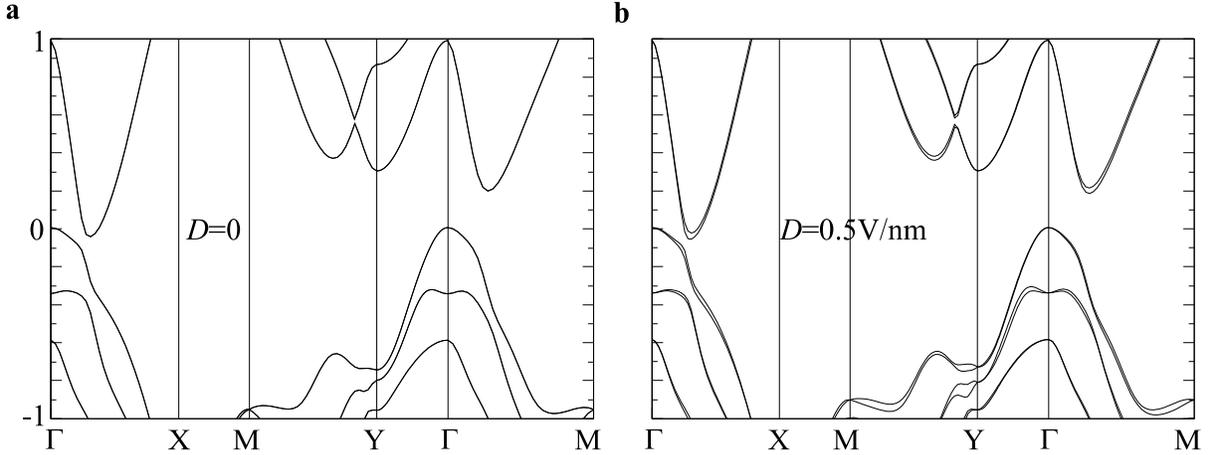

**Fig. S3. First-principle band structure in presence of displacement field $D$ without folding the BZ. a,** $D=0$. **b,** $D = 0.5$ V/nm. The energies are in units of eV. $D$ splits the spin degeneracy of the bands and also gives rise to larger Fermi surfaces for both electrons and holes simultaneously at charge neutrality.

## 2. Sample fabrication process

The detailed fabrication process of all three types of devices is summarized below and illustrated in Fig. S4.

### I. Transport device
1. Graphite and hexagonal boron nitride (hBN) exfoliation:
   a. Exfoliate graphite and hBN on cleaned $Si^{++}/SiO_2$ wafers.
   b. Search for proper flakes under optical microscope.
   c. Use AFM to identify the thickness and ensure the cleanness as well as the flatness of flakes.
2. Alignmark deposition:
   a. Dice undoped $Si/SiO_2$ (285 nm) wafer into appropriately sized pieces.
   b. Spin coat bilayer PMMA resist:
      i. 495PMMA A2, 2000 rpm for 1 min, and baked at 180 °C for 7 mins.
      ii. 950PMMA A2, 3000 rpm for 1 min, and baked at 180 °C for 3 mins.
   c. Pattern alignmarks by EBL (Raith e-Line) with 10kV acceleration voltage and 30 μm aperture size.
   d. Develop in cold bath (0 °C) with developer (IPA : DI water = 3:1 by volume) for 180 s.
   e. Deposit Ti (5 nm)/Au (50 nm) in Angstrom Engineering Nexdep e-beam evaporator.
   f. Lift off by immersing chip in acetone bath for 1 hour and then sonicate in another acetone bath for 5 mins. Transfer chip into dichloromethane (DCM) bath (soak for 1 hour) and acetone bath (soak for 3 hours) in successive order. Rinse with acetone and IPA, then dry with $N_2$ gas.
3. Bottom graphite and hBN transfer:
   a. Stack graphite/hBN and transfer on to alignmark chip with standard dry transfer technique.
   b. Remove polycarbonate (PC) by immersing chip in chloroform (1 hour), DCM (1 hour) and chloroform (3 hours) baths in successive order. Rinse with IPA and then dry with $N_2$ gas.
   c. Anneal in furnace with vacuum (~$10^{-4}$ torr) at 400 °C for 6 hours.
   d. Use AFM to find a clean usable area for contact electrodes.
4. Contact and gate electrodes creation:
   a. Spin coat bilayer PMMA resist (refer to **I**.2b).
   b. Pattern electrodes and gates using EBL with 10kV acceleration voltage and 30 μm aperture size.
   c. Pattern outer bonding pads and connections using EBL with 10kV acceleration voltage and 60 μm aperture size (high current mode).
   d. Develop in cold bath (0 °C) with developer (IPA : DI water = 3:1 by volume) for 180 s.
   e. Etch with reactive-ion etching (Oxford PlasmaPro 80 RIE) for 10-16 s.
   f. Deposit Ti (3 nm)/Pd (17 nm).
   g. Lift off by immersing chip in acetone (1 hour), DCM (1 hour) and acetone (3 hours) baths in successive order. Next, rinse with acetone and IPA, then dry with $N_2$ gas.
   h. Tip clean electrodes with AFM contact mode.
   i. Image the cleaned region with AFM tapping mode to ensure its cleanness and flatness.

5. Thin hBN transfer:
    a. Stack thin hBN (thickness: 2-5 nm) on top of contact electrodes with standard dry transfer technique.
    b. Remove PC by immersing chip in chloroform (1 hour), DCM (1 hour) and chloroform (3 hour) baths in successive order. Rinse with IPA and then dry with $N_2$ gas.
    c. Use AFM to ensure the cleanness and flatness of sample.
6. Thin hBN etching:
    a. Spin coat bilayer PMMA resist (refer to **I**.2b) on prepared bottom part from step 5.
    b. Pattern holes (~ 500nm × 500nm) using EBL on thin hBN at specific locations, where contact electrodes are underneath, with 10kV acceleration voltage and 30 μm aperture size.
    c. Etch holes on thin hBN with reactive-ion etching for certain time, depending on the thickness of thin hBN.
    d. Use AFM to ensure the contacts are well exposed.
    e. Remove PMMA resist by immersing chip in acetone (1 hour), DCM (1 hour) and acetone (3 hours) baths in successive order . Rinse with acetone and IPA, then dry with $N_2$ gas.
    f. Tip clean with AFM contact mode.
    g. Image the cleaned region with AFM tapping mode to ensure its cleanness and flatness.
    h. Immediately move the chip into an argon glovebox ($H_2O$ < 0.1 ppm, $O_2$ < 0.1ppm).
7. Top part stack transfer and final device assemble:
    a. Exfoliate, search and identify monolayer $WTe_2$ in glovebox with optical microscope.
    b. Stack $WTe_2$/hBN/graphite with standard dry transfer technique in glovebox.
    c. Transfer the stack onto designated area of prepared bottom part from step 6.
    d. Extract sample from glovebox, remove PC with chloroform (10 mins), DCM (1 min) and chloroform (10 mins). Then, rinse with IPA.
    e. Wire bond Al wires and quickly transfer sample into a fridge. The time of our final devices (with $WTe_2$ fully encapsulated) exposed to air is minimized to be less than 40 mins.

**II. vdW tunneling device**
1. Graphite and hBN exfoliation:
    a. Exfoliate graphite and hBN on cleaned $Si^{++}$/$SiO_2$ wafers of specific thickness. (for ultrathin hBN : 90nm $SiO_2$; for others : 285nm $SiO_2$)
    b. Search for proper flakes under optical microscope.
    c. Use AFM to identify the thickness and ensure the cleanness of flakes.
2. Alignmark deposition (refer to **I**.2).
3. Bottom vdW stack fabrication:
    a. Stack graphite/hBN/graphite with standard dry transfer technique in ambient condition.
    b. Remove PC by immersing chip in chloroform (1 hour), DCM (1 hour) and chloroform (3 hours) baths in successive order. Rinse with IPA and then dry with $N_2$ gas.
    c. Anneal in furnace with high vacuum (~$10^{-4}$ torr) at 450 °C for 3 hours.
    d. Use AFM to find a clean usable area.
4. Graphite tunneling fingers patterning:
    a. Spin coat bilayer PMMA resist (refer to **I**.2b).

b. Pattern graphite fingers using EBL with 30kV acceleration voltage and 10 μm aperture size.
   c. Develop in cold bath (0 °C) with developer (IPA :DI water = 3:1 by volume) for 5 s.
   d. Etch with reactive-ion etching for 38 s to remove excess graphite.
   e. Remove PMMA resist by immersing chip in acetone (1 hour), DCM (1 hour) and acetone (3 hours) baths in successive order. Rinse with IPA and then dry with $N_2$ gas.
5. Electrodes fabrication:
   a. Spin coat bilayer PMMA resist (refer to **I**.2b) on prepared bottom part from step 4.
   b. Pattern electrodes and gates using EBL with 10kV acceleration voltage and 30 μm aperture size.
   c. Pattern outer bonding pads and connections using EBL with 10kV acceleration voltage and 60 μm aperture size.
   d. Develop in a cold bath (0 °C) with developer (IPA : DI water = 3:1 by volume) for 180 s.
   e. Deposit Ti (3 nm)/Au (10 nm)/Pd (10 nm).
   f. Lift off by immersing chip in acetone (1 hour), DCM (1 hour) and acetone (3 hours) baths in successive order. Rinse with acetone and IPA, then dry with $N_2$ gas.
6. AFM Tip clean:
   a. Tip clean tunneling area (including graphite fingers and electrodes) with AFM contact mode.
   b. Image the cleaned region with AFM tapping mode to ensure its cleanness and flatness.
   c. Immediately move the chip into an argon glovebox ($H_2O$ < 0.1 ppm, $O_2$ < 0.1ppm).
7. Top vdW stack transfer and final device assemble:
   a. Exfoliate, search and identify monolayer $WTe_2$ in argon glovebox with optical microscope.
   b. Stack tunneling hBN/$WTe_2$/hBN/graphite with standard dry transfer technique in the glovebox.
   c. Transfer the stack onto designated area of prepared bottom part from step 6.
   d. Extract sample from the glovebox, remove PC with chloroform (2 baths, 10 mins each). Then, rinse with IPA.
   e. Wire bond Al wires and quickly transfer the sample into a fridge. The time of our final devices (with $WTe_2$ fully encapsulated) exposed to air is minimized to be less than 40 mins.

**III. STM Device**
1. Graphite and hBN flakes exfoliation (refer to **II**.1).
2. Prepatterned chip fabrication:
   a. Spin coat photoresist on the polished side of $Si^{++}$/$SiO_2$ (285 nm) chip: AZ1518, 4000 rpm for 40 s, and baked at 95 °C for 1 min.
   b. Dip the chip in 49% HF for 3 s.
   c. Deposit back gate contact Ti (3 nm)/Au (50 nm) on the unpolished side of the chip.
   d. Strip AZ1518 in 1165.
   e. Spin coat bilayer photoresist on polished side of the chip.
      i. LOR-3A, 4000 rpm for 40 s, and baked at 175 °C for 5 mins.
      ii. 1505, 4000 rpm for 40 s, and baked at 110 °C for 1 min.

f. Pattern contacts with photolithography on the chip using Heidelberg DWL66+.
   g. Develop the chip in 1165 at room temperature for 90 s.
   h. Descum the chip in TePla M4L Plasma Asher 5 mins.
   i. Deposit contact pattern Ti (3 nm)/Pd (20 nm) on the chip.
   j. Lift off by immersing chip in acetone (1 hour), DCM (1 hour) and acetone (3 hours) baths in successive order. Rinse with acetone and IPA, then dry with $N_2$ gas.
   k. Use AFM to ensure the cleanness and flatness of electrodes.
   l. Immediately move the chip into an argon glovebox ($H_2O$ < 0.1 ppm, $O_2$ < 0.1 ppm).
3. Device assemble:
   a. Exfoliate, search and identify monolayer $WTe_2$ in argon glovebox.
   b. Stack hBN/$WTe_2$/monolayer hBN with standard dry transfer technique in glovebox. (Remark: Bottom hBN is used to support $WTe_2$ to the same height of electrodes and hence minimize the local height fluctuation.)
   c. Transfer the stack onto designated area of the prepatterned chip.
   d. Extract sample from glovebox and remove PC with chloroform (2 baths, 10 mins each). Rinse with IPA.
   e. Tip clean the tunneling area with AFM contact mode.
   f. Wire bond Al wires and quickly transfer sample into STM UHV chamber. The time of the final device exposed to air is minimized to be less than 1 hour.

- **Summary of device parameters (Table S1)**

| Transport Device | D1 | D2 | D3 |
|---|---|---|---|
| Top Gate | Graphite ~ 7 nm | Graphite ~ 6 nm | $ZrTe_2$ (air sensitive) ~ 30 nm |
| Top hBN | 11 nm | ~ 8nm | 13.1 nm |
| Thin hBN | 4.8 nm | 2 nm | 4.8 nm |
| Bottom hBN | 9.2 nm | 6 nm | 15 nm |
| Bottom Graphite | 3.1 nm | 3 nm | 9 nm |
| Bulk $WTe_2$ | Flux growth | Vapor growth | Flux growth |

| vdW Tunneling Device | D4 | D5 | D6 |
|---|---|---|---|
| Top Graphite | 8.2 nm | 10.2 nm | 5 nm |
| Top hBN | 6.8 nm | 8 nm | 8.5 nm |
| Tunneling hBN | ~ 4 layers | ~ 4 layers | ~ 2 layers |
| Graphite Fingers | 4.5 nm | 5.2 nm | 8.7 nm |
| Bottom Dielectric | 17 nm hBN | 14 nm hBN | 285 nm $SiO_2$ |
| Bottom Gate | 2 nm Graphite | 3.6 nm Graphite | Si |

| STM Device | D7 |
|---|---|
| Top hBN | 1 layer |
| Bottom hBN | 13.5 nm |

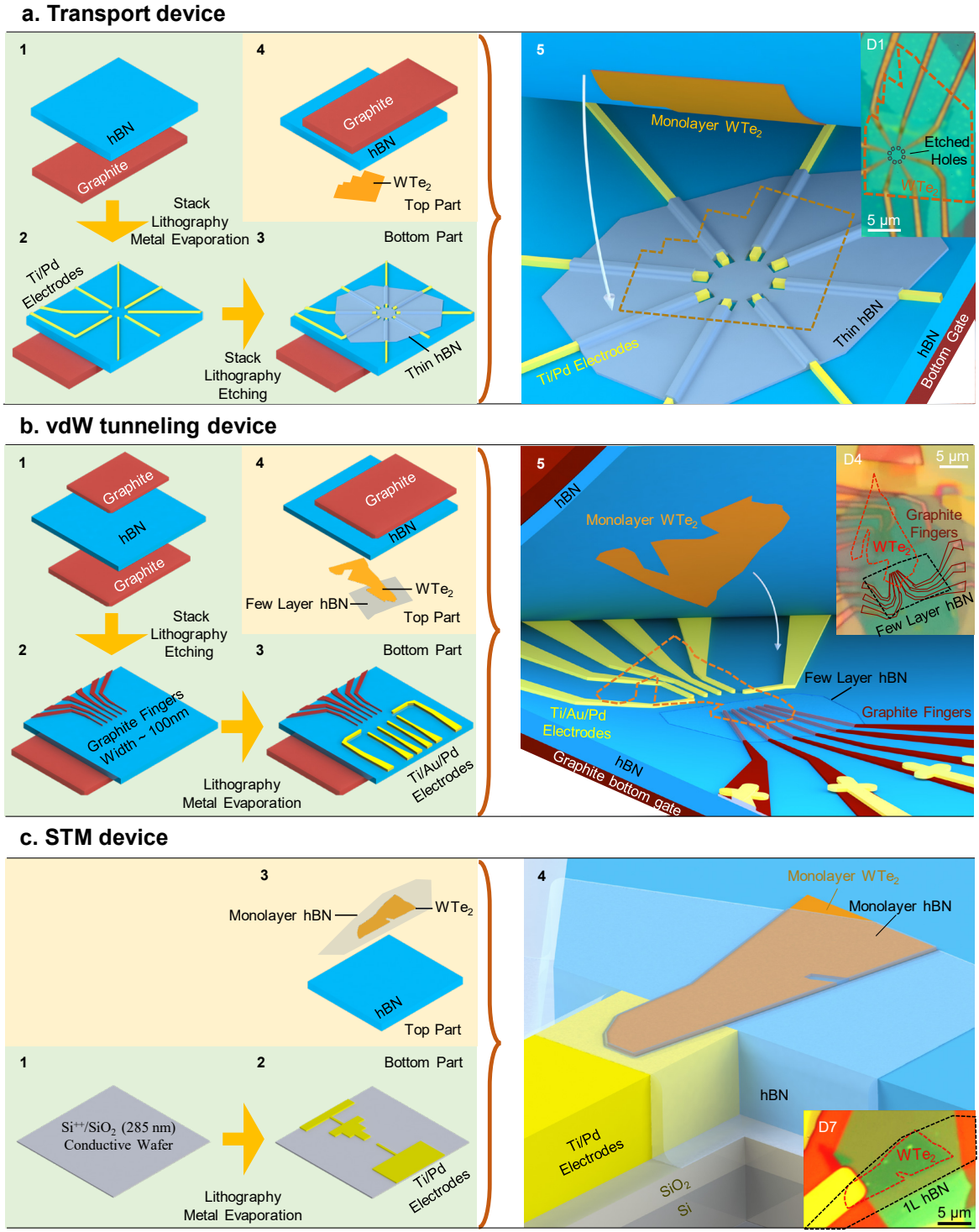

**Fig. S4. Device fabrication process.** Cartoon illustration of the fabrication process for **a**, transport device, **b**, vdW tunneling device and **c**, STM device. Insets in **a-c** are optical images of devices D1, D4 and D7, respectively.

## 3. Extracting $n_e$, $n_p$ and geometry factor α based on semiclassical model for D1

Transport in monolayer WTe$_2$ at high temperatures behaves like in a semimetal with contributions from both electrons and holes. In this section, we try to extract critical information of the system using transport data at 200 K by assuming that the semiclassical description is effective at such a high temperature. The semiclassical formula for longitudinal conductivity $\sigma_{xx}$ and Hall coefficient $R_H$ at moderate fields are

$$\sigma_{xx} = \gamma/R_{xx} = eu_h(n_e b + n_h), \tag{1}$$

$$\text{and } R_H = \alpha\, dR_{xy}^{as}/dB = (n_h - n_e b^2)/e(n_e b + n_h)^2. \tag{2}$$

Here $n_e$ ($n_h$) is the electron (hole) density, $u_e$ ($u_h$) is the corresponding mobility, $b = u_e/u_h$ is the ratio between electron and hole mobilities and $e$ is the elementary charge. $\gamma$ and α are the calibrating factors accounting for the real device geometry. Experimentally, $R_{xx}$ and $dR_{xy}^{as}/dB$ are measured (see below supplementary section 9) as a function of gate-induced density $n_g$, as shown in Fig. S5a and b for D1 at 200 K. Here $n_g$ is determined from gate capacitance, i.e., $n_g \equiv \varepsilon_r\varepsilon_0 V_{tg}/ed_{tg} + \varepsilon_r\varepsilon_0 V_{bg}/ed_{bg}$, where $\varepsilon_0$ is the vacuum permittivity, $\varepsilon_r$ is the relative dielectric constant of thin hBN and $d_{tg}$ ($d_{bg}$) is the thickness of hBN layer associated with the top (bottom) gate. $n_g$ is related to $n_e$ and $n_h$ in the system by

$$n_g = n_e - n_h + n_{g0}. \tag{3}$$

Here, $n_{g0}$ is an offset from zero $n_g$ due to unintentional doping during the fabrication process and can be determined by the location of CNP peak in $R_{xx}$ at low $T$. In D1, $n_{g0} = -3.291\times10^{11}$ cm$^{-2}$. At high temperatures, we assume that mobilities have weak gate dependence and introduce a new constant $\beta \equiv \gamma/eu_h$ to simplify the equations. Namely, in the analysis below we only assume that $u_h$ is gate-independent but still allow $u_e$ to vary. By solving equations (1) – (3), we obtain:

$$b = \frac{\beta R_{xx} - \alpha e\beta^2\, dR_{xy}^{as}/dB}{\beta R_{xx} + (n_g - n_{g0})R_{xx}^2}; \quad n_e = \frac{\beta/R_{xx} + n_g - n_{g0}}{b+1}; \quad n_h = n_e - n_g + n_{g0}.$$

Hence if we know the constants α and β, then $n_e$, $n_h$, and $b$ are all determined. We explore the effect of α and β in Fig. S5c-f. One can see that altering α will mostly affect the results on the electron side, while altering β mostly affects the hole side. If we apply the constraints that $n_e$ and $n_h$ must be positive and can only monotonically change with varying $n_g$, then the values of α and β become highly constrained. Proper values and the corresponding $n_e$, $n_h$ as well as $b$ extracted from the above formula are indicated by the colored curves in Fig. S5c-f. Dashed lines show typical results for unphysical α or β. We find that $7 < \alpha < 9$ and $8.5\times10^{16}$ Ω/cm$^2 < \beta < 9.5\times10^{16}$ Ω/cm$^2$ yield reasonably good results. Fig. S5g and h plot the results for α = 8 and β = 9 ×10$^{16}$ Ω/cm$^2$. The value of α is used to calibrate the Hall coefficient $R_H$ in our measurements.

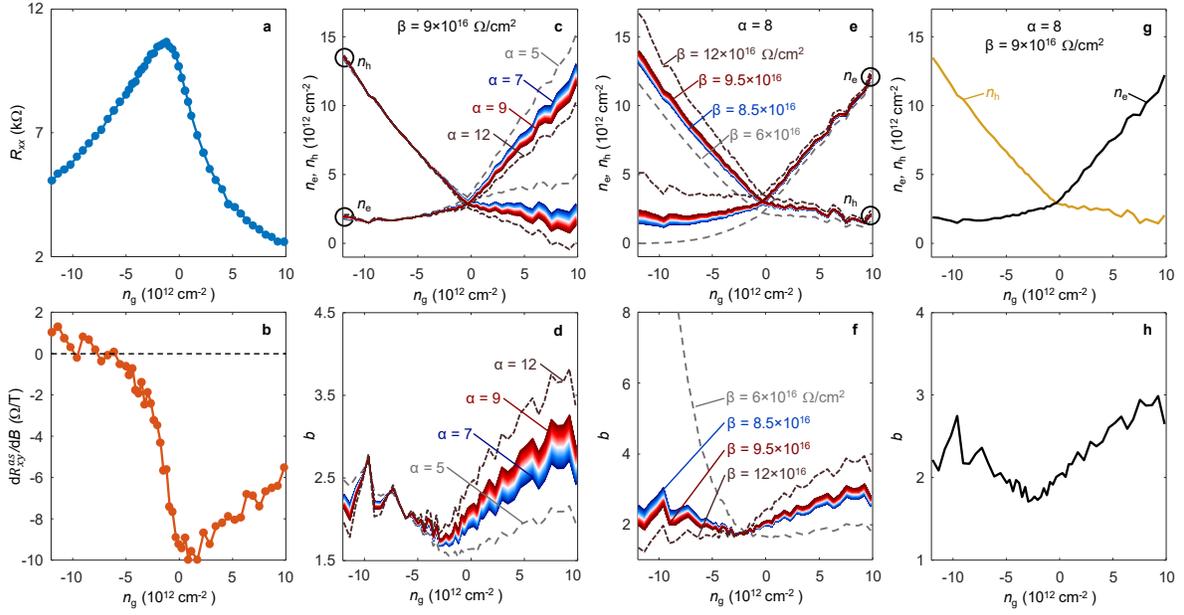

**Fig. S5. Semiclassical analysis of transport data at 200 K (D1). a,** Measured $R_{xx}$, and **b,** measured $dR_{xy}^{as}/dB$ as a function of $n_g$. **c** and **d,** The effect of the geometry factor α on the extracted densities $n_e$ and $n_h$ (**c**) as well as $b$ (**d**), based on the formula described in this section. Changing α mostly affects the curves in the electron side. Dashed lines indicate typical results of unphysical values of α, where negative or non-monotonically varying densities may be seen. The proper value of α are narrowly constrained, as indicated by the colored curves. **e** and **f,** The effect of β, which mostly affects the curves on the hole side. **g** and **h,** The results of $n_e$ and $n_h$ (**g**) as well as $b$ (**h**) for a proper choice of α and β.

## 4. Extracting $n_e$, $n_p$ and geometry factor α based on semiclassical model for D2

Following the same procedure described above, we perform the analysis for D2.

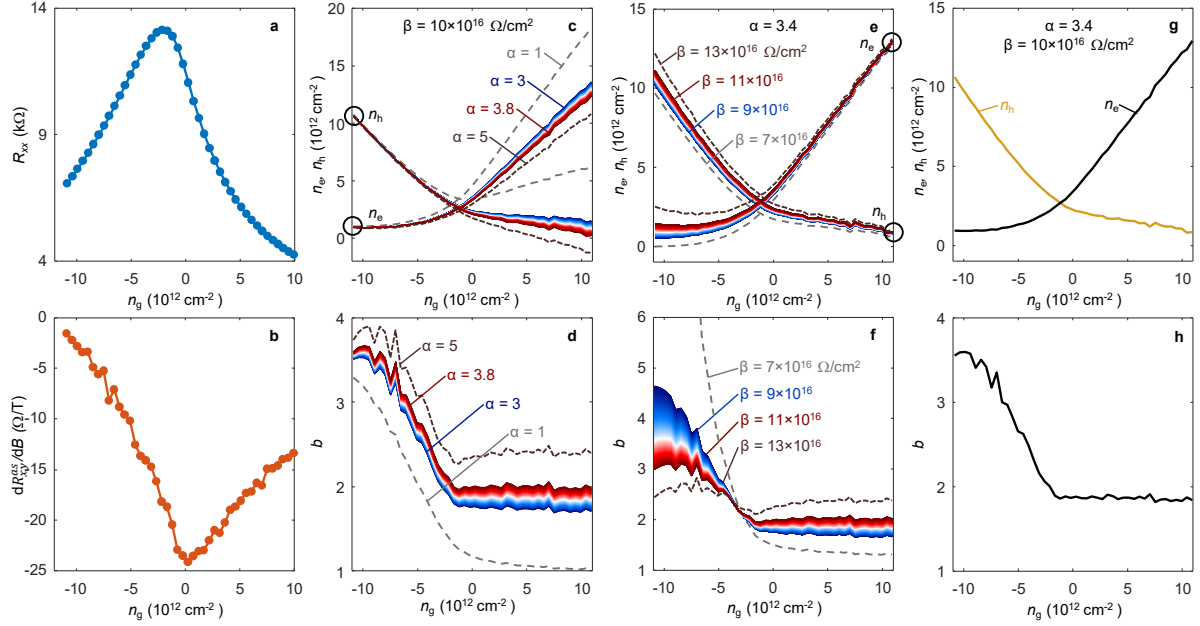

**Fig. S6. Semiclassical analysis of transport data at 200 K (D2). a,** Measured $R_{xx}$, and **b,** measured $dR_{xy}^{as}/dB$ as a function of $n_g$. **c** and **d,** The effect of geometry factor α on the extracted densities $n_e$ and $n_h$ (**c**) as well as $b$ (**d**). Dashed lines indicate typical results of unphysical values of α, where a negative or non-monotonically varying density may be seen. The proper value of α are narrowed constrained, as indicated by the colored curves. **e** and **f,** The effect of β, which mostly affects the curves in the hole side. **g** and **h,** The results of $n_e$ and $n_h$ (**g**) and $b$ (**h**) for a proper set of α and β, indicated in **g**. The smaller α compared to D1 is mostly due to the different choices of the Hall probes in the measurements.

## 5. RRR of flux-grown WTe₂ bulk

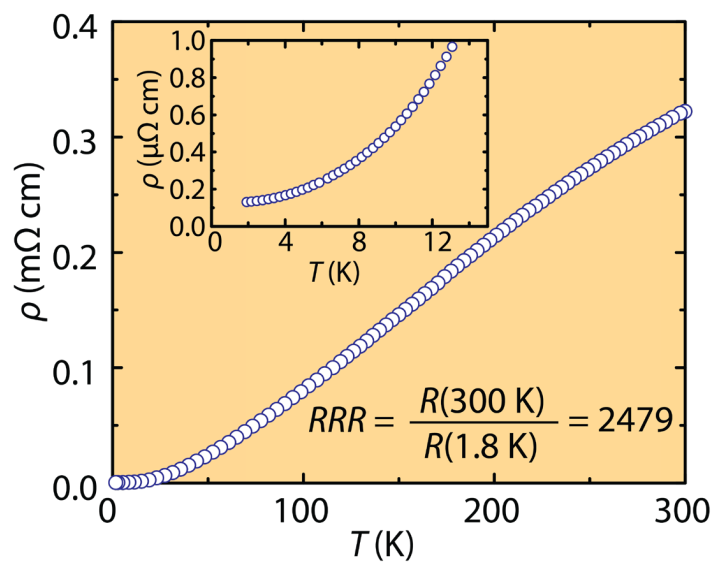

**Fig. S7. RRR of flux-grown WTe₂ bulk.** The resistivity of flux-grown WTe₂ bulk measured from 300K to 1.8 K. RRR is estimated to be 2479.

## 6. Gate-tuned resistance maps of D2 and D3

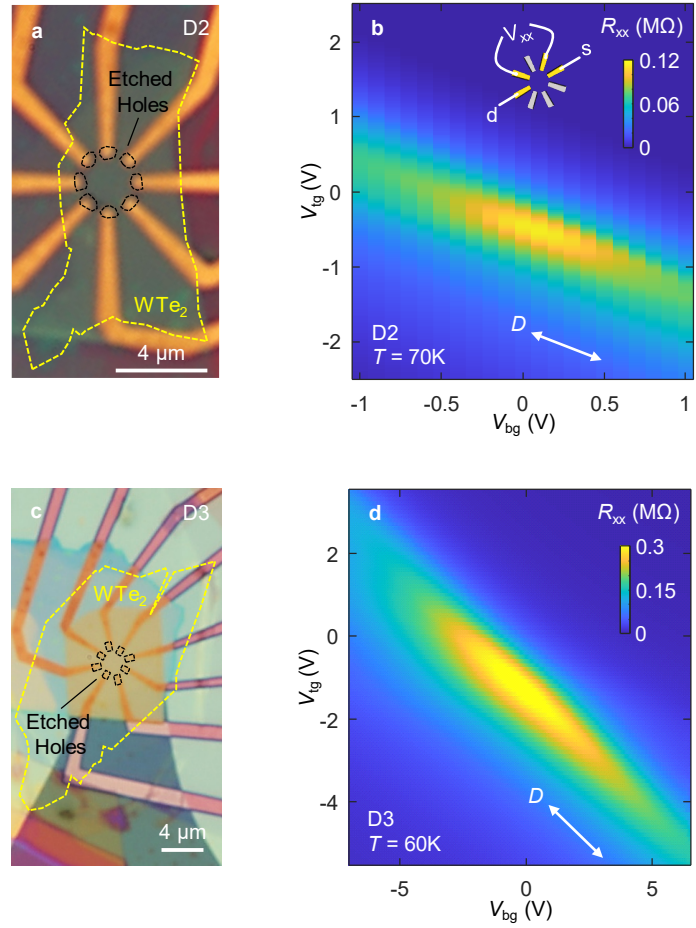

**Fig. S8. Device images and gate-dependent resistance of D2 and D3. a,** An optical image of D2. Dashed yellow lines indicate the edge of monolayer WTe$_2$. Black lines indicate the etched holes on the thin hBN layer, i.e., the contact regime. **b,** Resistance $R_{xx}$ tuned by top and bottom gates in D2, taken at 70 K. Inset shows the measurement configuration. **c** and **d,** The same plots for D3. Data taken at 60 K. Both devices show a strong insulating behavior along the charge neutrality line, despite the fact that the insulating peaks here are broader compared to D1. The strong effect of displacement field $D$ is also seen.

## 7. More discussion on the effect of disorders

As we have discussed in the main text, the effect of disorder can be evaluated by the broadness of the $R_{xx}$ peak, which reflects the inhomogeneity, and the offset of the peak from zero $n_g$, which reflects the unintentional doping induced by impurities. The differences between D1 and the other two devices (D2 and D3) on these aspects are clear (Fig. S9a, replotted from Fig. 2d). The peak of D1 is about one order of magnitude narrower than D2 and D3. Another way to look at the effect of disorders is in the conductance plot, as shown in Fig. S9b, where $G_{xx} \equiv 1/R_{xx}$ is plotted as a function of $n_g$. For all three curves, a linear behavior is well-approached in highly doped regime in both electron and hole sides. The slope of the linear trend reflects the mobility of the corresponding dominant carrier. Clearly, compared to D2 and D3, D1 (the black curve) exhibits higher mobilities for both electrons and holes, indicating much lower disorders/impurities in D1. At the same time, the resistivity at CNP in D1 is much higher, which implies that a cleaner sample hosts a stronger insulating state in the WTe$_2$ monolayer.

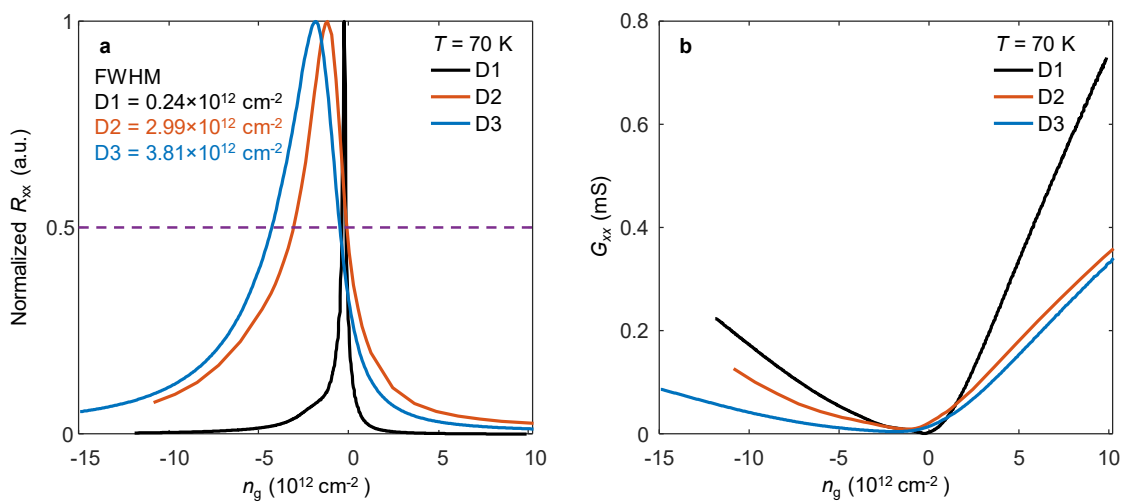

**Fig. S9. Resistance and conductance plots of D1-D3. a,** normalized $R_{xx}$, replotted from Fig. 2d. **b,** The same data, but plotted as $G_{xx} = 1/R_{xx}$.

## 8. More analysis on $R_{xx}$ curves

Here we analyze the temperature dependent resistance data in Fig. 2f based on both the standard activation formula, $R_{xx} \sim exp(\Delta/2k_BT)$, expected for a conventional band insulator and the Efros–Shklovskii (ES) variable-range hopping formula, $R_{xx} \sim exp[(T_{ES}/T)^{1/2}]$, expected for a Coulomb gap[5] for a 2D localized system. Here $k_B$ is the Boltzmann constant, $\Delta$ is the activation gap and $T_{ES} \sim e^2/\kappa\xi$ is a characteristic temperature determined by the localization strength $\xi$ and dielectric constant $\kappa$ of the enviroment. We plot the data using the Arrhenius plot with an axis of $T^{-1}$ shown in Fig. S10a and a similar plot but with an axis of $T^{-1/2}$ in Fig. S10b. Under both plots, the $R_{xx}$ curves display the two-stage behavior in the temperature range of 200 K ~ 20 K. Note that $R_{xx}$ at $D \sim 0.15$ V/m roughly shows a linear behavior in both plots due to the small temperature window (much less than a decade variation). The sample is too insulating so we cannot access values at lower temperatures reliably for that curve. In general, $R_{xx}$ has a strong dependence on $D$ at low $T$. In the Coulomp gap scenario, the strong $D$ dependence implies that $D$ field produces a dramatic effect on $\xi$ or $\kappa$. However, as the localization strength $\xi$ characterizes the tunneling probability between nearest disordered sites and dielectric constant $\kappa$ is mostly affected by the hBN substrate, we are not aware of any reliable process or prior examples for $\xi$ or $\kappa$ to be so sensitive to the $D$ field. In the scenario of a conventional band gap, the observation would imply closing of the single-particle gap by $D$, which may be reasonable as $D$ breaks inversion symmetry and introduces spin splitting at the band edges. However, our experiments on the Hall effect and the tunneling spectroscopy rule out the band insulator scenario. Overall, within the decade of temperature range, the bended curves show that the $R_{xx}$ data obeys neither the activation formula nor the ES variable hopping formula.

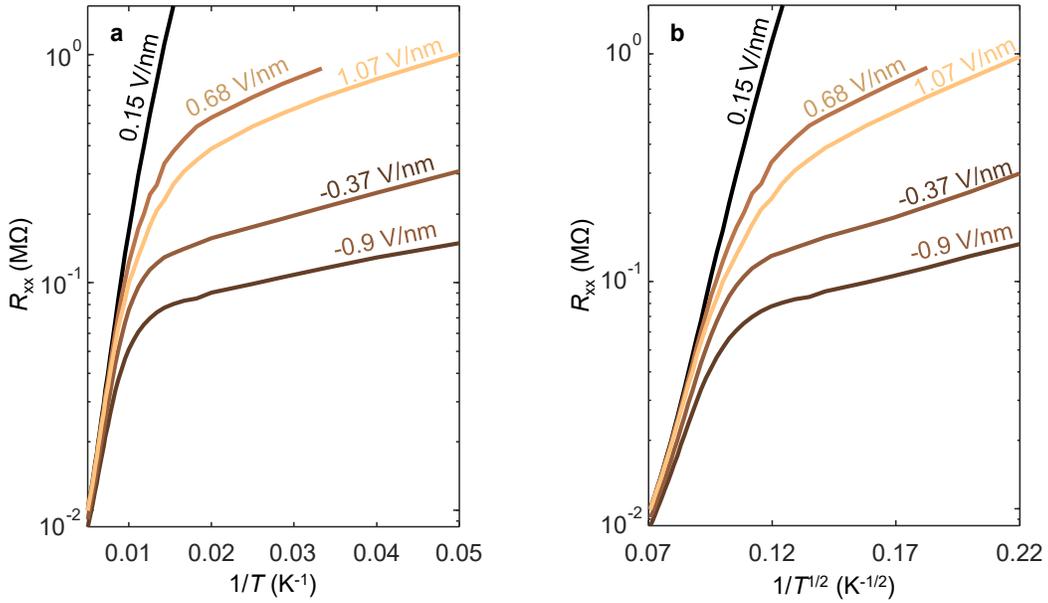

**Fig. S10. $R_{xx}$ vs $1/T$ & $R_{xx}$ vs $1/T^{1/2}$. a,** The $R_{xx}$ data replotted from Fig. 2f. **b,** The same data plotted as a function of $T^{-1/2}$.

# 9. Extracting the Hall coefficient $R_H$ from the raw $R_{xy}$ data

As mentioned in the main text, $R_H = \alpha \, dR_{xy}^{as}/dB$, where $R_{xy}^{as}$ is the asymmetric component of the measured $R_{xy}$ and $\alpha$ is a geometry factor determined from supplementary section 3 & 4. We extract $R_{xy}^{as}$ from $R_{xy}$ using $R_{xy}^{as}(B) = (R_{xy}(B) - R_{xy}(-B))/2$. Typical curves of $R_{xy}^{as}$ used in the main text are presented along with their corresponding raw $R_{xy}$ in Fig.S11 below. We find that $R_{xy}^{as}$ is well approximated as a linear function of $B$ in our measurement ($B$ up to 4.5 T), i.e., $R_H$ is well defined.

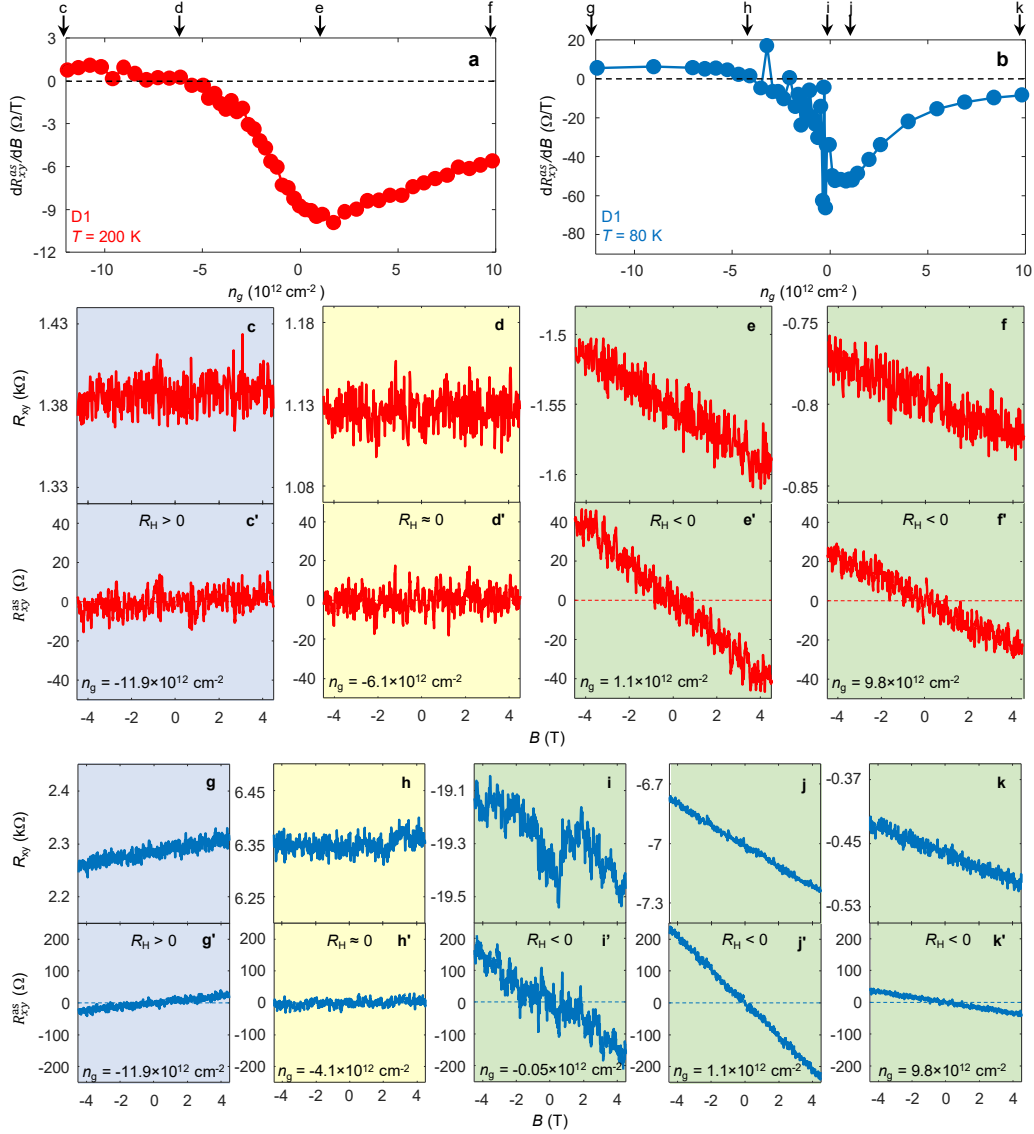

**Fig. S11. Extracting the Hall coefficient. a and b,** Extracted $dR_{xy}^{as}/dB$ for D1 at 200 K and 80 K respectively (the same data used in Fig. 3b). **c-f,** Typical raw $R_{xy}$ data and **c'-f',** the asymmetric component $R_{xy}^{as}$ of data points in **a**, indicated by the arrows. **g-k,** Typical raw $R_{xy}$ data and **g'-k',** the asymmetric component $R_{xy}^{as}$ of selected data points in **b**, indicated by the arrows.

## 10. Characterizing the vdW tunneling devices and the STM device

In Fig. S12, we present a collection of typical tunneling spectra taken at a large bias window (> 800 mV) for both vdW tunneling devices (D4 & D5) and the STM device (D7). The data also agree well with previous reports on STM spectra taken on $WTe_2$ monolayer grown on graphene[6,7], including the characteristic high-energy peak at ~ -630 mV. While consistent results are seen in all curves in Fig. S12, there are also quantitative differences between them, such as the sharpness of the high energy peak, which we attribute to the different qualities of hBN tunneling barrier and the monolayer $WTe_2$ used in our devices. Summary of all device parameters are shown in Table S1. Their low-energy behavior (< 100 meV) with in-situ electrostatic gating are summarized in Fig. 4, Fig. S14 and S15.

In our vdW tunneling devices, there are additional features located near ± 70 mV, where the spectra develop a sudden drop in the $dI/dV$ amplitude (Fig. S12b). This effect has been well studied previously in vdW tunneling devices using hBN as the tunneling barrier and has been understood as the effect of phonons[8].

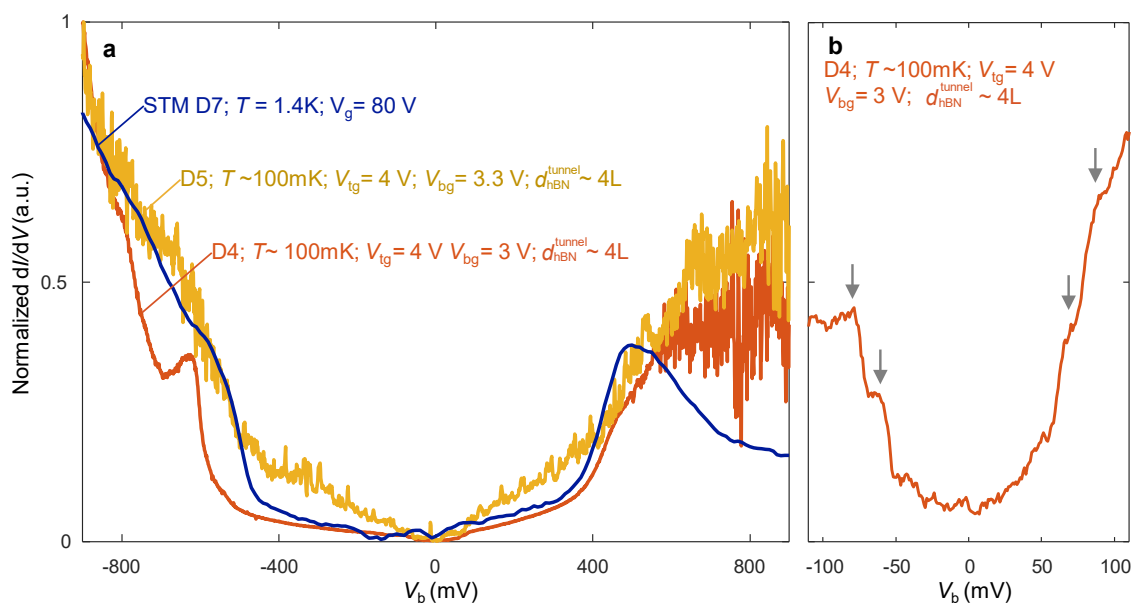

**Fig. S12. Large bias tunneling spectra for different devices. a,** Typical $dI/dV$ spectra taken from vdW tunneling devices (D4 & D5) and the STM device (D7). The conditions for taking the spectra are indicated next to the curves. In vdW devices, the bottom gates are used to enhance the conductivity of the monolayer flakes away from the tunneling junctions. **b,** Phonon characteristics shown in the vdW tunneling spectrum, exhibited as a sudden change in signal near ± 70 mV, indicated by the gray arrows. Details of this feature are studied in ref[8].

## 11. Characterizing the tunneling gap

Figure S13 displays the typical tunneling spectra taken at the insulating state near the CNP of monolayer $WTe_2$ (vdW device D4 and STM device D7). Fully depleted U-shape gaps are observed in both devices. The data are plotted in both linear scale (Fig. S13 a & b) and log scale (Fig. S13 c & d). In the linear scale plots, the locations where we start to see a clear $dI/dV$ signal above noise level are identified by the dashed red lines. A better characterization of the gap is performed based on the log scale plots, where we fit the rise of the $dI/dV$ signal away from zero and extract a tunneling gap of a size ~ 47 meV for the vdW device (D4) and ~ 91 meV for the STM device (D7), respectively.

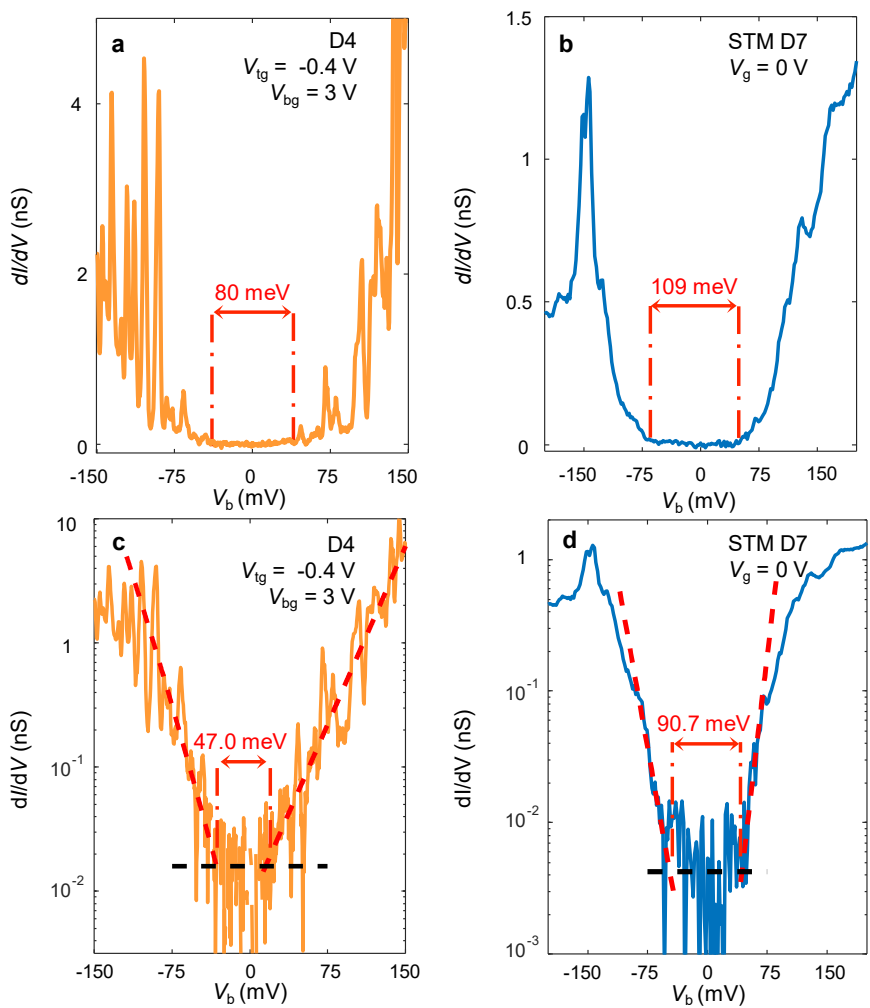

**Fig. S13. Characterizing the tunneling gap. a,** The low bias tunneling spectrum taken at the insulating state in vdW tunneling device D4, corresponding to a line cut in Fig. 4d. **b,** A corresponding line cut in Fig. 4f at the insulating state. **c** and **d,** The same data plotted in log scale.

## 12. Additional STM data on D7

Fig. S14a shows a typical STM topography of our sample surface, which is dominated by a stripe-like pattern, expected for WTe$_2$ monolayer lattice. The monolayer lattice is a distorted 1T-type structure, where every other tellurium atom is lifted out of the top lattice plane (Fig. S14b). The STM tip is most sensitive to these elevated atomic rows, resulting in the anisotropic surface topography observed in STM experiments. This indicates that electrons from the tip can tunnel through the hBN capping layer, allowing for measuring the underlying WTe$_2$. We note that, compared to previous studies, the topography in our experiment does not resolve the individual atom in the monolayer, which can probably be attributed to the presence of the protective hBN monolayer that inhibits atomic resolution imaging. Fig. S14 c-e present the gate-tuned d$I$/d$V$ spectra taken at three different locations in D7. The exact pattern in the maps differs, which reflects different disorders and screening environments at different locations. Nevertheless, all maps consistently demonstrate that the insulating state arises due to a correlated gap at the Fermi surface. We note that the tunneling spectra taken in the metallic regime are smooth while substantial charging lines develop in the insulating phase (for both STM and vdW tunneling devices), probably because of localized charges appearing due to poor charge dissipation in the insulator.

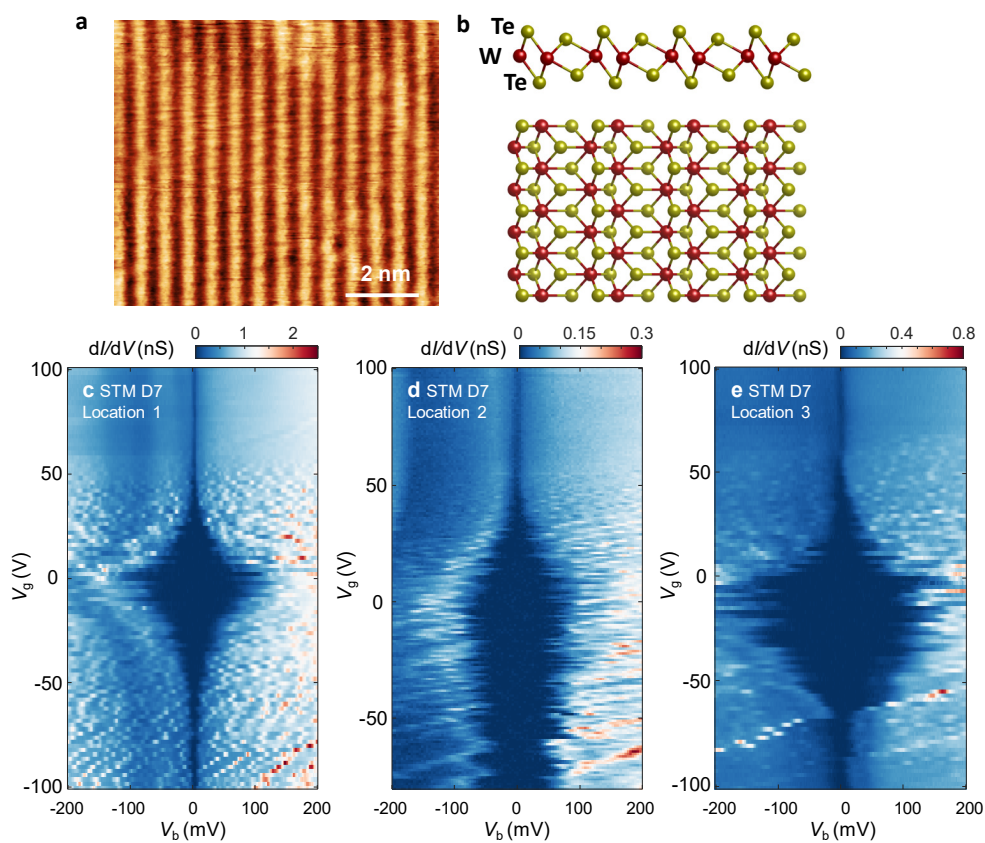

**Fig. S14. Additional STM data. a,** A typical STM topography taken from D7. $V_g$ = 80 V, $V_b$ = -800 mV, and the current was set at 10 pA. **b,** Lattice structure of WTe$_2$ monolayer. Upper: side view. Lower: top view. **c-e,** Gate-tuned d$I$/d$V$ spectra taken at three different locations on the STM device (D7). Data are taken at 1.4 K.

## 13. Summary of tunneling data taken from additional vdW tunneling devices (D5 & D6)

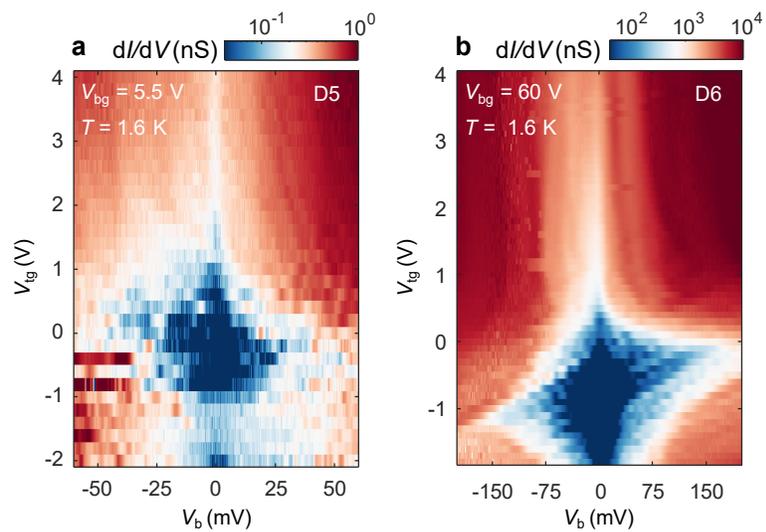

**Fig. S15. Gate-tuned d$I$/d$V$ spectra of additional vdW tunneling devices (D5 & D6, a,b).** Device parameters are summarized in table S1. $T$ = 1.6 K.

## 14. Explanation on the tunneling gap pinned at the Fermi energy

Here we provide a possible explanation on why the observed excitonic gap is pinned at Fermi energy when the doping in the $WTe_2$ monolayer varies. As illurstrated in Fig. S16, due to the presence of three pockets (two electron pockets and one hole pocket) in the $WTe_2$ monolayer, the nesting of the pockets is not perfect even at charge neturality. In the excitonic insulator phase, the nesting is optimized by self-adujsting the necessary q vector (see supplementary section 1 and Fig. S1b), which leads to an excitonic gap at the Fermi surface (Fig. S16a). With doping, the same optimization process occurs, but a different q vector may be chosen, again resulting in an excitonic gap at the Fermi surface with additional Fermi surfaces occupied by the extra carriers (Fig. S16b). In Fig. S1b, our susceptibility calculation indeed suggests that the q vector is not sharply peaked at a specific value and may be adjusted over a range based on the details of the Fermi pockets. As a result, in the gate-tuned tunneling experiments, we expect to observe a gap pinned at the Fermi energy. Away from the CNP, the extra carriers introduced to the sample will reduce the size of the gap and fill in the gap (they may be localized in real devices). This is consistent with our observatons in tunneling expierments. Note that we aim at a conceptual illurstration here; a precise understanding of the system requires more sophisticated modeling.

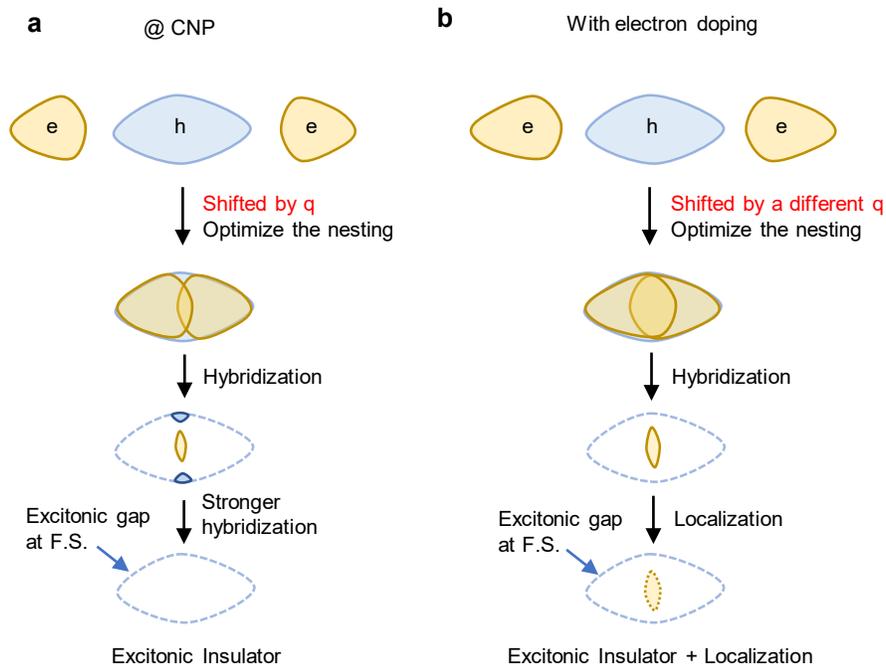

**Fig. S16. Conceptual demonstration of the excitonic gap pinned at Fermi energy in $WTe_2$ monolayer**. **a,** Fermi surface nesting and hybridization for producing the excitonic insulator at CNP. **b,** The gap opening process for the case of electron doping. There is no perfect nesting for both cases. The necessary q vector for producing the band overlap will be adjusted for different doping to optimize the nesting.